\def\conff{0}

\ifnum\conff=0
\documentclass[12pt]{article}
\fi

\ifnum\conff=1
\documentclass{llncs}
\fi
\usepackage[active]{srcltx}
\usepackage{times}
\usepackage{url}
\usepackage{comment}
\usepackage[english]{babel}
\usepackage{array}

\usepackage{algorithm}
\usepackage{subfig}
\usepackage{amsmath}
\usepackage{amssymb}
\usepackage{wrapfig}
\usepackage{graphicx}
\graphicspath{{./Figs/}}

\usepackage{paralist}

\ifnum\conff=0
\usepackage[a4paper]{geometry}
\usepackage{setspace}%\singlespacing
\fi

\ifnum\conff=0
\usepackage{amsthm}
 \newtheorem{theorem}{Theorem}

\newtheorem{claim}[theorem]{Claim}

\fi

\newenvironment{proof sketch}[1]{\noindent {\emph{Proof sketch of #1:}}}{\hfill \qed}

\newcommand{\eqdf}{\triangleq}

\newcommand{\SINR}{\text{\sc{sinr}}}
\newcommand{\SNR}{\text{\sc{snr}}}
\newcommand{\PER}{\text{\sc{per}}}
\newcommand{\pps}{\text{\emph{pps}}}
\newcommand{\hops}{\text{\emph{hops}}}
\newcommand{\mf}{\text{\emph{mf}}}
\newcommand{\MCS}{\text{\sc{mcs}}}
\newcommand{\Ein}{E_{\text{\emph{in}}}}
\newcommand{\Eout}{E_{\text{\emph{out}}}}

\newcommand{\algA}{\textsc{MF-I-S}}

\newcommand{\algB}{\textsc{ShortP}}
\newcommand{\algBS}{\textsc{ShortP-S}}
\newcommand{\algC}{\textsc{MF-I}}
\newcommand{\algD}{\textsc{MF}}
\newcommand{\algE}{\textsc{MF-S}}

\newcommand{\algS}{\algB}

\begin{document}
%\begin{titlepage}
\title{Real-Time Video Streaming in Multi-hop Wireless Static Ad Hoc Networks}

\ifnum\conff=1
\author{%
Guy Even
\and
Yaniv Fais
\and
Moti Medina
\and Shimon (Moni) Shahar
\and Alexander Zadorojniy}

\institute{School of Electrical Engineering,
Tel-Aviv Univ., Tel-Aviv 69978, Israel.
\protect\url{{guy,yanivfai,medinamo,moni,sasha}@eng.tau.ac.il}}
\fi

\ifnum\conff=0
\author{%
Guy Even\thanks{School of Electrical Engineering,
Tel-Aviv Univ., Tel-Aviv 69978, Israel.
\protect\url{{guy,yanivfai,medinamo,moni,sasha}@eng.tau.ac.il}}
\and
Yaniv Fais$^*$
\and
Moti Medina$^*$
\and Shimon (Moni) Shahar$^*$
\and Alexander Zadorojniy$^*$ }
\fi

\date{}

\maketitle

%%% ----------------------------------------------------------------------

\begin{abstract}
  We deal with the problem of streaming multiple video streams between
  pairs of nodes in a multi-hop wireless ad hoc network.  The nodes
  are static, know their locations, and are synchronized (via GPS).  We introduce a new interference model that
  uses variable interference radiuses. We present an algorithm for
  computing a frequency assignment and a schedule whose goal is to
  maximize throughput over all the video streams.  In addition, we
  developed a localized flow-control mechanism to stabilize the queue
  lengths.

  We simulated traffic scheduled by the algorithm using OMNET++/MixiM
  (i.e., physical SINR interference model with 802.11g) to test
  whether the computed throughput is achieved. The results of the
  simulation show that the computed solution is \SINR-feasible and
  achieves predictable stable throughputs.
\end{abstract}

\begin{comment}
  \paragraph{ Keywords:}
\end{comment}

%\end{titlepage}
\section{Introduction}

We address the problem of routing real-time video streams in static ad
hoc wireless networks.  Our goal is to develop and implement an
efficient algorithm and test it in a realistic physical model.  Many
works have been published on the topic of multi-hop routing in
wireless networks including real-time video streaming (see
\cite{setton2005cross,khan2006application,shan2005cross,van2005cross}).
In these works it is acknowledged that cross layer algorithms are
required to utilize the capacity of the network.  These papers
evaluate specific algorithms and scenarios using approximate models
for wireless network, and thus the question of developing integrated
realistic solutions remains open. In particular, a solution must
address a combination of specifications including: maximize
throughput, fairness, minimize delay, stability of throughput,
stability of queue lengths in intermediate nodes, bounded number of
lost packets, and predictability.

One of the main issues in wireless networks is how to model
interferences. In the communication community, one uses the
signal-to-interference-plus-noise ratio (SINR) to determine if a
received signal is decoded without an
error~\cite{gallager1968information}. On the other hand, the
algorithms community has used the graph model (or protocol model) to
model feasible communication
patterns~\cite{jain2005impact,alicherry2005joint11}. For the graph
model, multi-hop routing algorithms with a constant approximation
ratio have been
developed~\cite{kumar2005algorithmic,alicherry2005joint11,buragohain2007improved,wan2009multiflows}.
In fact, Wan~\cite{wan2009multiflows} even presents a (theoretical)
PTAS for the problem. On the other hand, to date approximation
algorithms for throughput maximization in the SINR model do not have a
constant approximation ratio. For example, in~\cite{ChafekarCapacity},
the approximation ratio is logarithmic in the ratio between the
longest link and the shortest link (for uniform transmission powers),
and in~\cite{EMM11} the approximation ratio is logarithmic in the
number of nodes (for the linear power model).

The study of wireless algorithms in the SINR model has been motivated
by its realistic appeal. In fact, it has been argued that the
performance of graph based algorithms is inferior to algorithms in the
SINR model~\cite{goussevskaia2007complexity}.
In~\cite{moscibroda2006protocol,moscibroda2006topology} a logarithmic
ratio between the throughput in the SINR model and the throughput in
the graph model is presented.  A closer look at studies that compare
the interference models and algorithms for these models shows only a
constant gap if the ratios of the max-to-min power and max-to-min
distance are constant.  In~\cite{gupta2000capacity} the same
asymptotic throughput is obtained in both models with respect to
random instances. In ~\cite{moscibroda2006protocol}, the example only
gives a constant ratio if the the power ratio and the distance ratio
are constant.  In~\cite{ChafekarCapacity}, an example with a constant
gap is presented for constant uniform power.
In~\cite{behzad2004performance}, the theorems do not utilize the
ability to increase the interference radius or to apply collision
avoidance methods used in the 802.11 MAC.

The questions we study in this paper are as follows.
\begin{enumerate}[(i)]
\item How much of the traffic computed by a graph model based routing
  algorithm can be routed in realistic scenarios with constant
  max-to-min powers and constant max-to-min distances? Namely, does
  the approximate nature of the graph model lead to useful solutions?
\item How to integrate a graph-model based routing-algorithm in a
  system that supports real-time video streaming? Such a system must
  combine goals such as: fairness, predictable throughput, few lost
  packets, bounded intermediate queues, reasonable and steady
  end-to-end delay.
\end{enumerate}

\paragraph{Previous Work.}  The necessity of cross layer designs has been
recognized for satisfying the special characteristics of real-time
video streaming over wireless
networks~\cite{shan2005cross,setton2005cross,khan2006application}.  We
continue this line of work.

\begin{comment}
  The WiFi technology builds on the 802.11 protocol, where in this
  paper we used the 802.11g standard. The exact details of the
  communication, such as RTS/CTS frames, and hold off time are
  described in detail in ~\cite{gast2005802}. The standard has 8
  Modulation Coding Schemes (\MCS) that define the data-rate, coding
  and modulation. The \MCS s are ordered from a rate of 6Mbps for
  \MCS\ $0$ to rate of 54Mbps for \MCS\ $7$.  The higher the
  data-rate, the higher the required signal-to-noise ratio (\SNR).
  Adapting the \MCS\ is one of the ways to decrease the error rate in
  case of change in the \SNR
  ~\cite{kamerman1997wavelan,holland2001rate}.
\end{comment}

The multi-hop routing problem for ad hoc networks was investigated
thoroughly.
\begin{comment}
  While our emphasis is on supporting video streams (i.e., high
  throughout and low end-to-end delay), many works focus on
  distributed algorithms for packet routing in the case of mobile
  nodes~\cite{johnson1996dynamic,broch1998performance,perkins2002ad}.
\end{comment}
One of the commonly used heuristics for routing is based on finding
paths with maximum bottlenecks, namely, paths for which the edge with
the lowest capacity is maximum~\cite{draves2004routing}.  We used this
algorithm in our benchmarks (we call it \algS). A different approach
for the routing problem is based on solving a linear program.  In
~\cite{kumar2005algorithmic,jain2005impact,alicherry2005joint11},
routing algorithms in the graph model are designed, analyzed, and
simulated.  One drawback
in~\cite{kumar2005algorithmic,jain2005impact,alicherry2005joint11} is
that the simulations were run also in the graph model and not in the
physical model.  Wan~\cite{wan2009multiflows} pointed out various
errors in previous algorithms and presented a new linear program that
corrects the problem. He proved that: (i) there is a
$23$-approximation algorithm based on the linear program, and (ii)
there is a polynomial time scheme (PTAS) for the problem. However,
this PTAS is not practical.  Namely, the PTAS requires solving a
linear program that might not be solved by LP-solvers even for
moderate sized networks.

Chafekar et al.~\cite{ChafekarCapacity,chafekarPhD} considered routing
algorithms in the SINR model. The approximation ratio of their
algorithm with uniform power assignments is logarithmic in the ratio
between the longest link and the shortest link.
\begin{comment}
  From a high level view, their algorithm is similar to the algorithms
  in
  \cite{kumar2005algorithmic,alicherry2005joint11,buragohain2007improved}.
  Namely, an LP is formulated for a multi-commodity problem that
  includes also interference constraints. The solution of the LP is
  transformed into a schedule by applying greedy coloring. The
  analysis in~\cite{ChafekarCapacity} is rather involved since it
  needs to translate interferences caused by a transmitters to
  interferences accumulated by each receiver.
\end{comment}
Their communication model does not include \textsc{ack} packets.
Hence, interference is caused only by the sender and not by the
receiver.  In~\cite{chafekarPhD}, simulations are described in the
physical model, but these simulations do not use the 802.11 MAC (i.e.,
no \textsc{rts, cts, ack} packets are used).

Three scenarios are simulated in~\cite{chafekarPhD}: random network,
grid, and a realistic road-traffic network. The path loss exponent in
the simulations is $\alpha=6$, which is considered rather high and fit
for indoor environments (but not open air environments). In the grid
scenario, the nodes are located $10$ meters apart (both horizontally
and vertically), and the communication range also equals $10$ meters.
Thus, communication is possible only between closest neighbors.
Interference caused by a node that is located a diagonal away (i.e.,
$10\sqrt{2}$ meters away) is $8$ times smaller, and interference
caused by a node $20$ meters away is $64$ times smaller. Thus,
interferences in this setting fade very quickly, justifying a small
gap between the \SINR\ model and the graph based protocol model.
Another aspect in the simulation of~\cite{chafekarPhD} is that routing
is limited to single paths (i.e., no splitting). It is not clear if
this is an implementation issue or a result of the simulated
instances.

\paragraph{Special Characteristics of Real-time Video Streaming.}
%\label{sec:video streaming}
Streaming of real-time video in a multi-hop ad hoc network is a
challenging task with unique characteristics.
\begin{enumerate}[(1)]
%  \item Video coding (e.g., H.264/MPEG) creates a bursty stream since
%    I-frames are much longer than the other frames. One could smoothen
%    the stream at the cost of incurring a extra delay of a second
%    (i.e., the time between I-frames).
\item End-to-end delay in streaming of real-time video should be as
  small as possible.  We assume that a delay of $1$-$2$ seconds is
  tolerable if the video has to travel across $10$ hops.  One
  implication of this constraint is that end-to-end acknowledgments and TCP are
  not an option.
\item Unless erasure codes are employed, loosing even a small fraction
  of the packets incurs an intolerable degradation in the video
  quality. We assume that video has acceptable quality if less that
  $0.5\%$ of the packets are dropped. In wireless networks, each link
  can have a \PER\ of $1\%$-$5\%$. Thus, after $10$ hops, one is left
  with intolerable erasures.  On the other hand, erasure codes incur an
  extra end-to-end delay since they need to accumulate data for a
  block before encoding can take place. This means that relying on
  WiFi acknowledgments and retransmit capabilities can be useful to
  avoid packet drops if the \PER\ is small.
\item A useful feature in video coding is the ability to adjust the
  compressed bit-rate. This means that the video encoder can be
  continuously controlled to generate a video stream of a requested
  bit rate. We rely on this feature in our flow control algorithm.
  This feature separates video streaming from other applications such
  as FTP.
\end{enumerate}

\paragraph{Our Contributions.}
\begin{enumerate}
\item We do not modify the 802.11g MAC. This approach has two
  advantages. First, we do not bypass the wireless NIC and its
  collision avoidance features. Hence, even if the algorithm suggests
  a schedule with interferences, these interferences are resolved by
  the MAC. Second, the network can support limited additional traffic
  that is not routed or scheduled by the algorithm (i.e. messages for
  controlling the network). We choose the 802.11g because of its
  popularity in laptops and mobile devices.
\item Simulation in the physical model. The simulation is in a
  standard 802.11g setting using OMNET++/MixiM (see
  Sec.~\ref{sec:exp}).  In this setting, all WiFi frames are
  transmitted (i.e., RTS,CTS, packet, ACK), and interferences between
  frames are analyzed using the SINR-model, and taking into account
  the Modulation Coding Schemes (\MCS).
\item We introduce new interference constraints that constitute an
  intermediate model between the physical \SINR-model and the graph
  based protocol model (see Sec.~\ref{sec:models}). The interference
  set of a link is a function of the signal-to-noise ratio of the link
  and the \MCS\ of the link.  As the signal-to-noise ratio (without
  interferences) of a link is closer to the \SINR-threshold, the
  interference set grows, so that \SINR\ is not in the ``waterfall''
  region of the \PER\ function.\footnote{The packet-error-rate (\PER)
    is a function of the SINR. This function increases very steeply in
    the neighborhood of the critical threshold $\beta$. This
    phenomenon is referred to as the ``waterfall'' region of the \PER\
    function.  }  One advantage of this new interference model is that
  it is easy to formulate interference constraints in the linear
  program formulation (see Sec.~\ref{sec:LP} in the Appendix).
\item We formulate the problem of minimizing end-to-end delay incurred
  by a schedule that supports a given multi-flow. We developed and
  implemented a scheduling algorithm that addresses this problem of
  reducing end-to-end delays while supporting a similar throughput
  (see Sec~\ref{sec:sched}).
  In~\cite{kumar2005algorithmic,alicherry2005joint11,buragohain2007improved}
  the effect of the schedule on the delay is not mentioned.
\item We developed and implemented a flow control algorithm that
  stabilizes the queue lengths and controls the data-rate along the
  links.  This flow control algorithm is executed locally by the
  nodes.
\item We evaluated the performance of the proposed algorithm with
  respect to video streaming. In particular, we measured the
  throughput, end-to-end delay, fraction of dropped packets, queue
  lengths, and the stability of these parameters.
\end{enumerate}

\paragraph{Techniques.}
Following~\cite{kumar2005algorithmic,alicherry2005joint11,buragohain2007improved,ChafekarCapacity,EMM11},
we formulate an LP, and apply greedy coloring to obtain a schedule.
Interestingly, the greedy coloring incurs high end-to-end-delays, so
we developed a path-peeling scheduler that trades delay for
throughput. Stability is maintained by a flow control algorithm that
monitors flow through incoming and outgoing links, and continuously
balances the two. This method utilizes the ability of video encoders
to adjust the compressed bit-rate.

\begin{comment}
  \paragraph{Organization.}

  In Appendix~\ref{sec:video streaming} we discuss the special
  characteristics of real-time video streaming.
\end{comment}

\section{Problem Definition}\label{sec:problem}
\paragraph{Setting.}
We consider a WiFi 802.11g static ad hoc network with $3$ non-interfering radio
channels with the assumptions:
\begin{inparaenum}[(i)]
\item Single radio: each node has a single wireless network interface
  controller (WNIC).
\item Each node is equipped with a GPS so that it knows
  its location and the nodes are synchronized.
\item The WNICs support quick synchronized hops between frequency
  channels.
\item Isotropic antennas.
\item We also assume that the nodes have already joined the network
  and that there is at least one node (i.e., center node) that holds
  full information about the network (i.e., nodes and locations).
  Accumulating this information can be done in a distributed
  low-bandwidth fashion after building a spanning
  tree~\cite{awerbuch1987optimal}.
\end{inparaenum}

\paragraph{Problem Definition.}
The input to the algorithm consists of:
\begin{enumerate}
\item A set $V$ of $n$ nodes in the plane. A transceiver is located in
  each node.
\item A set of $k$ video stream requests $\{r_i\}_{i=1}^k$. Each
  stream request is a triple $r_i\eqdf (a_i,b_i,d^*_i)$, where $a_i$ is
  the source (e.g., camera) of the stream, $b_i$ is the destination,
  and $d^*_i$ is the required data-rate.
\end{enumerate}
Ideally, we would like to satisfy all the requests, namely, for each
video stream $r_i$, route packets using multi-hops from $a_i$ to $b_i$.
We assume that there is a path in the network between each
source-destination pair (otherwise, the request is rejected).

Let $d_i$ denote the data-rate achieved for the $i$th stream.  The
service ratio $\rho_i$ of the $i$th demand is defined by $\rho_i \eqdf
d_i/d^*_i$.  Our goal is to maximize the minimum service ratio,
namely, $\max \min_i \rho_i$.

Additional performance measures are:
\begin{inparaenum}[(i)]
\item End-to-end delay - this is the time it takes a packet to reach
  its destination. We are interested in reducing the maximum delay
  (among the packets that are delivered) since the video is real-time.
  In addition, the maximum delay determines the size of the jitter
  buffer in the receiving side.
\item Number of dropped packets. Queue management may drop packets. A dropped packet
never reaches its destination.
\item Queue lengths in intermediate nodes tell us how much memory
  should be allocated and also give an indication of the delay per hop.
\end{inparaenum}

\section{Preliminaries}
\label{sec:prelim}

\subsection{Interference Models}\label{sec:models}

\paragraph{Bidirectional interference.}
The delivery of a message in the WiFi MAC requires transmission of
frames by both sides (e.g., RTS and packet are transmitted by the
sender, CTS and ACK are transmitted by the receiver). Hence,
interferences can be caused also by frames transmitted by a the
receiving side.

\paragraph{The SINR model.}
The SINR model, also called the physical
interference model, defines successful communication as follows.  Let
$d_{u,v}$ denote the distance between nodes $u$ and $v$.  Suppose a subset
$S_t\subseteq V$ of the nodes are transmitting simultaneously in the
same frequency channel as $u$.  The signal-to-interference-plus-noise
ratio (SINR) for the reception by $v\in V\setminus S_t$ of the signal
transmitted by $u\in S_t$ in the presence of the transmitters $S_t$ is
defined by
\[
\SINR(u,v,S_t) \eqdf \frac{P/d_{u,v}^{\alpha}}{N+\sum_{x\in
    S_t\setminus\{u\}} P/d_{x,v}^{\alpha}}\:.
\]
Each transmitter can use one of several modulation coding schemes (\MCS).
The message transmitted by $u$ in an \MCS\ $m$ is successfully received by
$v$ if $\SINR(u,v,S_t)\geq \beta_m$, where $\beta_m$ is the minimum
SINR-threshold for the \MCS\ $m$.

\paragraph{Protocol model.}
The protocol model, also called the graph model, is specified by two
radii:
\begin{inparaenum}[(i)]
\item A communication distance $r$.
\item An interference distance $R$.
\end{inparaenum}
The rule for successful communication between two nodes $u$ and $v$ is
that $v$ receives the message from $u$ if $d_{u,v}<r$ and every other node
$x$ that transmits at the same time satisfies $d_{x,v}> R$.  In this
model, a communication graph is defined over the nodes.  Two nodes are
linked by an edge if their distance is less than the communication
distance $r$.

Since the WiFi MAC requires transmission by both sides, an
\emph{interference} is defined between two links $(u,v)$ and $(u',v')$
if $\min\{d_{u,u'},d_{u,v'}, d_{v,u'}, d_{v,v'}\} < R$.  We say that a
subset $L$ of links is \emph{non-interfering} if no two links in $L$
interfere.  In the protocol model, a \emph{schedule} is a sequence
$\{L_i\}_i$ of subsets of non-interfering links.

\paragraph{Our new model.}
The new model is an intermediate model between the SINR model and the
protocol model. The idea is that, as the \SNR\ of a link grows, the
link can tolerate more interference. Hence, the interference distance
is not fixed.

Consider a pair $(u,v)$ of nodes and an \MCS\ $m$.
The triple $(u,v,m)$ is a \emph{link} in the new model if
$\SINR(u,v,\emptyset) \geq \beta_m$.

Since both sides of a link transmit and receive, the interference set
of a link must take into account interferences caused by other
transmissions both in the receiver and the sender. However, the frames
sent by the receiving side are in \MCS\ $0$, therefore, reception of
these frames depends on the $\SINR$-threshold $\beta_0$.

The interference set $V_{u,v,m}$ of the link $e=(u,v,m)$ is defined by
  \begin{align*}
    V_{u,v,m} \eqdf \left\{ x \in V \setminus\{u\}\right. &\left | \SINR(u,v,\{x\}) < \mu  \cdot \beta_m \text{ or } \right. \\
    & \left.\SINR(v,u,\{x\}) < \mu \cdot \beta_0\right\}.
\end{align*}
The motivation for this definition is that transmissions of nodes in $V_{u,v,m}$ interfere with
the reception of $v$ by $u$, or vice versa.
The choice of $\mu=1.585$ gives us a margin of $2$dB above the
\SINR-threshold.  This margin keeps the $\SINR$ above the threshold
due to interferences caused by transmitters not in $S_{u,v,m}$.

We also define the interfering set of edges with respect to the link $e=(u,v,m)$.
\begin{align*}
  I_{u,v,m} \eqdf &\{ e'=(u',v',m') \mid \{u',v'\} \cap (V_{u,v,m} \cup V_{v,u,m})
 \neq \emptyset\}\\& \setminus \{(u,v,m)\}.
\end{align*}
The interference set $I_{u,v,m}$ contains a link $e'$ if either
endpoint of $e'$ interferes with reception at the endpoints $u$ or
$v$.

\begin{comment}
  \begin{align*}
    S_{u,v,m} &\eqdf \left\{ x \in V \left | \frac{1}{6}\cdot
        \frac{P}{d_{x,v}^{\alpha}} > \left( \frac{P}{\beta\cdot
            d_{u,v}^{\alpha}} - N \right) \right.  \right\}
  \end{align*}
The justification for the definition of the interference set is summarized in the following claim.
\begin{claim}
Suppose $S_t$ is ``feasible''.
If $S_t \cap S_{u,v,m}=\emptyset$ then
$\SINR(u,v,S_t) \geq \SINR (u,v,\emptyset) \cdot \gamma.$
\end{claim}
\end{comment}

\paragraph{Notation.}
Let $u$ and $v$ denote nodes and $m$ denote an \MCS.  A link is a
triple $(u,v,m)$ such that $\SINR(u,v,\emptyset)\geq \beta_m$.  This
definition implies that there can be multiple parallel links between
$u$ and $v$, each with a different \MCS.
We denote the set of links by $E$. The set $\Eout(v)$ (resp.
$\Ein(v)$) denotes the set of links that emanate from (resp. enter)
$v$. Let $E(v)$ denote the set of links $\Ein(v)\cup \Eout(v)$.
For a link $e=(u,v,m)$, let $\MCS(e)=m$, i.e., the \MCS\ $m$ of the link $e$.

\section{Algorithm \algA}\label{sec:outline}
\subsection{Networks Governed by Time-Slotted Frequency Tables}\label{sec:govern}
Two tables govern the communication in the network.  The first table
$A$ is a time-slotted frequency table. The dimensions of $A$ are
$F\times T$, where $F$ denotes the number of frequency channels and
$T$ denotes the number of time slots.  There is one row for each
frequency channel and one column for each time slot. (In our
implementation we used $F=3$ and $T=200$).  The table $A$ determines a
periodic schedule.  The second table is a multi-flow table $\mf$. The
dimensions of $\mf$ are $|E|\times k$ (recall that $k$ equals the
number of video streams).  The entry $\mf(e,s)$ specifies the number
of packets-per-period that should be delivered along link $e$ for
stream $s$.

Each table entry $A[j,t]$ is a subset of links,
i.e., $A[j,t]\subseteq E$.  The table governs communication in the
sense that, in slot $t'$, the links in $A[j,t' \pmod T]$ try to deliver packets
using frequency channel $j$.

We use $A[\cdot,t]$ to denote the set of links $\cup_{j\in F}
A[j,t]$.  Since we assume that each node is equipped with a single
radio, it follows that two links that share an endpoint cannot be
active in the same time slot.  Hence, for every node $v$, $E(v) \cap
A[\cdot ,t]$ may contain at most one link.

A time-slotted frequency tables schedules active links as listed in
Algorithm TX-RX in Appendix~\ref{sec:algs}.  Each node $v$
executes Algorithm TX-RX$(v)$ locally.  Since $E(v) \cap A[\cdot ,t]$
may contain at most one link, a node $v$ is either a receiver, a
sender, or inactive in each time slot.

\subsection{Algorithm Specification}
The input to the routing algorithm is specified in
Sec.~\ref{sec:problem}.
The output consists of two parts:
\begin{inparaenum}[(i)]
\item a time-slotted frequency table $A$, and
\item a multi-flow $\mf(e,s)$, for every link $e$ and stream $1\leq
  s\leq k$.
\end{inparaenum}
We note that the units of flow are packets-per-period. The period
equals $T \cdot \sigma$, where $\sigma$ is the duration of a time
slot, and $T$ equals the number of time-slots in a period.

The multi-flow $\mf(e,s)$ determines the routing and the throughout of
each stream. The role of the frequency/time-slot table $A$ and the
multi-flow tables is to specify a periodic schedule that determines
which links are active in which time slots (see Sec.~\ref{sec:govern}).

Although we use fixed length packets (e.g., 2KB), the \MCS\ of a link
determines the amount of time required for completing the delivery of
a packet.  This means, that within one time slot, multiple packets may
be delivered along a single link.  Let $\pps(e)$ denote the number of
packets-per-slot that can be delivered along $e$.  Namely, node $u$
can transmit at most $\pps(e)$ packets to node $v$ along link
$e=(u,v,m)$ in one time-slot. Note that the value of $\pps(e)$ is a function of
the \MCS\ of the link $e$.

\medskip \noindent We say that table $A$ \emph{supports} the flow
$\mf$ if the following properties hold:
\begin{enumerate}
\item Every entry $A[j,t]$ in the table is a set of non-interfering
  links. Thus, the links in $A[j,t]$
  may be active simultaneously.
\item The data-rates $\mf(e,s)$ are supported by the table. Namely,
  \begin{align}\label{eq:support}
    \sum_{s=1}^k \mf(e,s) \leq |\{A[j,t] : e\in A[j,t]\}|\cdot \pps(e)\:.
  \end{align}
\end{enumerate}

\subsection{Algorithm Description} \label{sec:sched}
Algorithm \algA\ consists of two parts: (i)~computation of a
multi-commodity flow with conflict constraints, and (ii)~scheduling of
the multi-commodity flow in a time-slotted frequency table.
We elaborate on each of these parts.

\paragraph{Multi-commodity flow with conflict constraints.}
We formulate the problem of routing and scheduling the video streams
by a linear program (LP).  A similar LP is used
in~\cite{kumar2005algorithmic,alicherry2005joint11,buragohain2007improved} with respect to
the graph model.  We use our new interference model
for the interference constraints.

The variables $f^j_i(e)$ of the LP signify the amount of
flow along link $e$ in frequency channel $j$ for stream $i$.
The full LP  appears in Appendix~\ref{sec:LP}.
Let $f^j(e) \eqdf\sum_{i=1}^k
  f^j_i(e)$, namely, $f^j(e)$ is the flow in frequency $j$ along link
  $e$.  Let $c(e)\eqdf T \cdot \pps(e)$ denote the number
  of packets-per-period that can be delivered along the link $e$.

\medskip
\noindent
We elaborate on two main features of the LP:
\begin{enumerate}
\item The conflict constraints.  The ratio $f^j(e)/c(e)$ equals the
  fraction of the time that the link $e$ is active in transmission in
  frequency $j$. Since each node is equipped with a single WNIC,
  transmissions emanating or entering the same node may not occur
  simultaneously (in any frequency). In addition, the links in $I_e$
  may not transmit in frequency $j$ whenever $e$ is transmitting in
  frequency $j$. Thus, the conflict constraint is formulated as follows. For
  every link $e=(u,v,m)\in E$, and for each frequency $j\in[1..3]$:
  \begin{align*}
        \frac{f^j(e)}{c(e)} + \sum_{j'\neq j} \sum_{e'\in E(u)\cup E(v)}
    \frac{f^{j'}(e')}{c(e')} + \sum_{e' \in I_{e}}
    \frac{f^j(e')}{c(e')} & \leq 1\:. &
% \forall e=(u,v,m)\in E,\forall    j\in[1..3] \label{eq:conf}
  \end{align*}
\item Max-Min throughput.
For each requested stream $r_i$, we define the supply ratio $\rho_i$
to be the ratio between the flow allocated to the $i$'th stream and
the demand $d^*_i$ of the stream.  The objective of the LP is to
maximize $\min_i \rho_i$.  A secondary objective is to maximize the
total throughput.
\end{enumerate}

\paragraph{Scheduling of the multi-commodity flow in a time-slotted
  frequency table.}
In the scheduling step we are given the multi-commodity flows
$f^j_i(e)$.  The task is to allocate entries in a
time-slotted frequency table $A$ that supports these flows.

We first determine how many time-slots should be allocated for
$f^j(e)$, for each link $e$ and each frequency channel $j$.  Similarly to Eq.~\ref{eq:support},
\begin{align*}
  |\{ t\in[1..T] :e\in A[j,t]\}| \cdot \pps(e) &\geq f^j(e)
\end{align*}
Hence,
\begin{align}\label{eq:req}
  |\{ t\in[1..T] :e\in A[j,t]\}| \geq \left\lceil \frac{f^j(e)}{\pps(e)} \right\rceil.
  \end{align}
\paragraph{The greedy scheduler.}
  The simplest way to assign flows to the table $A$ is by applying a
  greedy algorithm (similar to greedy coloring).  The greedy algorithm scans the links and
  frequency channels, one by one, and assigns $\ell(e,j)$ slots to
  each link $e$ and frequency channel $j$.  Based
  on~\cite{alicherry2005joint11,kumar2004end,buragohain2007improved},
  the interference constraints in Eq.~\ref{eq:conf} imply that the
  greedy algorithm succeeds in this assignment provided that
\begin{align}\label{eq:guar}
 \ell(e,j) = \left\lfloor \frac{f^j(e)}{\pps(e)} \right\rfloor.
  \end{align}
  The issue of dealing with this rounding problem (i.e., the
  difference between the round-down and the round-up in Eqs.~\ref{eq:req} and~\ref{eq:guar}) is discussed
  in~\cite{wan2009multiflows}, where it is pointed out that routing
  all the flow requires a super exponential period $T$. Such a period
  is obviously not practical; the computation of the table takes too
  long, the table is too long to be broadcast to all nodes, and
  the schedule will incur huge delays.

  We show that the rounding problem is not an important issue both
  theoretically and in practice.  Since each flow $f_i$ can be
  decomposed into at most $|E|$ flow paths, it follows that the values
  of $\{f^j_i(e)\}_{e\in E, j\in F}$ can be ``rounded'' so that at
  most $|E|\cdot \max_e \{\pps(e)\}$ packets are lost per period.
  Note that this lost flow can be made negligible by increasing the
  period $T$.  As $T$ increases, the amount of flow per period tends
  to infinity, and hence, the lost flow is negligible. In our
  experiments~\ref{sec:exp}, we used a period of $T=200$ time slots,
  with a duration of $5ms$ per slot. The greedy scheduler was able to
  schedule almost all the flow in all the instances we considered.
  The multi-flow table is set so that $\mf(e,s)$ equals the amount of
  flow from $f_s(e)$ that the scheduler successfully assigned.

 The greedy scheduler incurred a delay roughly of
  one period per hop. The reason is that it schedules all the
  receptions to a node before the transmissions from the node. To
  avoid this delay, we designed a new scheduler, described below.

\paragraph{The path-peeling scheduler.}
The path peeling scheduler tries to reduce the time that an incoming
packet waits till it is forwarded to the next node. This is achieved
as follows.
\begin{enumerate}
\item Decomposes each flow $f_i$ into flow paths such that the flow
  along each path equals the bottleneck, i.e., the minimum $\pps(e)$ along the path.  Let
  $\{f_i(p)\}_{p\in {\cal{P}}(i)}$ denote this decomposition.
\item While not all the flow is scheduled,
  \begin{enumerate}
  \item \label{item:robin} For $i=1$ to $k$ do:
  \item If ${\cal P}(i)\neq \emptyset$, then schedule a path $p\in {\cal P}(i)$ and remove $p$ from ${\cal P}(i)$.
  \end{enumerate}
\end{enumerate}
The scheduling of a flow path $p\in {\cal P}(i)$ tries to schedule the links in
$p$ one after the other (cyclically) to reduce the time a packet needs to wait in each node
along $p$. The scheduling simply scans the links in $p$ in the order
along $p$, and finds the first feasible time slot (in cyclic order) for
each link $e\in p$.

We point out that in Line~\ref{item:robin}, we schedule one path from
each stream to maintain fairness in allocation and delays. On the average, each stream
suffers from the same ``fragmentation'' problems in the table $A$.

In our experiments, the path-peeling scheduler succeeded in scheduling
70\% of the flow. The advantage, compare to the greedy
scheduler, is that delays are significantly reduced.

\section{Flow Control}\label{sec:flow control}
The multi-flow table computed by the algorithm determines the number of packets $\mf(e,s)$
that should be sent along each link $e$ for stream $s$ during each period.
Each node $v$ monitors the following information for each link $e \in \Eout (v)$.
\begin{enumerate}
\item $P(e,s,t)$ - the number of packets belonging to stream $s$ sent
  along the link $e$ during the period $t$.
\item $P^+(e,s,t)$ - the maximum number of packets belonging to stream
  $s$ that can be sent along the link $e$ during the period $t$. Note
  that $P^+(e,s,t)\geq P(e,s,t)$; inequality may happen if the queue
  $Q(e,s)$ is empty when a packet is scheduled to be transmitted along
  the link $e$. Note that if $e$ is not planned to deliver packets of
  stream $s$, then $P^+(e,s,t)=0$.
\end{enumerate}
We remark that a node $v$ can also monitor $P(e,s,t)$ for a link $e\in
\Ein(v)$.  However, the value $P^+(e,s,t)$ for a link $e\in \Ein(v)$
must be sent to $v$ (e.g., by appending it to one of the delivered
packets).

The Flow-Control algorithm is executed locally by all the nodes in the
network. Let $e=(u,v,m)$ denote a link from $u$ to $v$, and let $s$
denote a stream. Each node executes a separate instance per stream.
In the end of each period $t$, each node $u$ ``forwards'' the value of
$P^+(e,s,t)$ to node $v$.  In addition, in the end of each period $t$,
node $v$ sends ``backwards'' the value $R(e,s)$ to $u$. The value
$R(e,s)$ specifies the number of packets from stream $s$ that $v$ is
willing to receive along the link $e$ in the next period
$t+1$.

\begin{algorithm}
  \caption{Flow-Control$(v,s)$ - a local algorithm for managing the
    local queue and requested incoming rate at node $v$ for stream
    $s$.}
\label{alg:FC}
  \begin{enumerate}
  \item Initialize: for all $e\in \Ein(v)$, $R(e,s)\gets \mf(e,s)$.
  \item For $t=1$ to $\infty$ do
    \begin{enumerate}
    \item Measure $P(e,s,t)$ for every $e\in E(v)$, and $P^+(e,s,t)$ for every $e\in\Eout(v)$.
    \item Receive $P^+(e,s,t)$ for every $e\in\Ein(v)$, and $R(e,s)$ for every $e\in \Eout(v)$.
    \item \label{line:Rin}
$R_{in} \gets \min\{\sum_{e\in\Eout(v)} R(e,s),$\\
$ \sum_{e\in\Eout(v)} P^+(e,s,t), \sum_{e\in\Ein(v)} P^+(e,s,t),
            \}$.
          \item \label{line:Res}
For every $e\in\Ein(v)$: $R(e,s) \gets R_{in} \cdot
            \frac{P^+(e,s,t)}{\sum_{e'\in\Ein(v) P^+(e',s,t)}}$.
            \item Drop oldest packets from $Q(v,s)$, if needed, so
              that $|Q(v,s)|\leq R_{in}$.
    \end{enumerate}
  \end{enumerate}
\end{algorithm}

The Flow-Control algorithm is listed as Algorithm~\ref{alg:FC}. It
equalizes the incoming and outgoing packet-rates in intermediate nodes
as follows.  The requested packet-rate $R(e,s)$ is initialized to be
the value $\mf(e,s)$ derived from the table.  The Flow-Control
algorithm is activated in the end of each period.  It uses the values
$P(e,s,t)$ and $P^+(e,s,t)$ for every link $e$ incident to $v$.  Some
of these values are computed locally and some sent by the neighbors.
The incoming packet-rate $R_{in}$ is computed in line~\ref{line:Rin},
and is divided among the incoming links in line~\ref{line:Res}. Excess
packets in the queue $Q(v,s)$ are dropped so that the number of
packets in $Q(v,s)$ is at most $R_{in}$. The rational is that, in the
next period, at most $R_{in}$ packets will be delivered, and hence,
excess packets might as well be dropped.

We now elaborate on the boundary cases of the flow-control for a
source $a_s$ and a destination $b_s$ of stream $s$.  The destination
$b_s$ simply sends a fixed request for each incoming link $e\in \Ein(b_s)$, i.e.,
$R(e,s)\gets \mf(e,s)$.  The source $a_s$, does not execute
line~\ref{line:Res}; instead, it sets the packet-rate of the video
encoder to $R_{in}$.

\section{Experimental Results}
%In this section we summarize the main experimental results.
%An elaborated discussion of the experimental results appear in Appendix~\ref{sec:exres}.

\subsection{General Setting}\label{sec:exp}

\paragraph{WiFi parameters.}  In the benchmarks that use the scheduler,
each node has a single 802.11g WNIC.  In the benchmarks that do not
use the scheduler, each node has three 802.11g WNICs.  The reason is
that, in absence of the scheduler, a node does not know to which
frequency channel to tune in each moment.

 Each WNICs transmits in one of three
non-overlapping frequencies.  All WNICs transmit at a fixed power
($100mw$). The path loss exponent is $\alpha=4.1$. The noise figure
is $N=-100dBm$. The maximum communication is roughly $150m$ (in \MCS\
$0$). An interference $250m$ away can cause a decrease in the \SINR\
of roughly $1dB$. We used fixed size packets with a payload of $2KB$.
Thus, a video stream with a $1$Mbps generates $64$ packets per second.

\paragraph{Software tools.}  We used Coin-OR CLP to solve the linear
programs.  We implemented the scheduler in C++.  The simulation was
implemented using OMNET++/MixiM. Therefore, the simulation is done in
the physical model taking into account path loss, multiple
interferences, partial interference between frames, and all the
details of the 802.11g protocol.

\paragraph{Algorithm parameters.} We used $T=200$ time slots in the
time-slotted frequency table.  Each time slot has a duration of $5$ms.

\paragraph{Implementation details.}
The following simplifications we made in the implementation.
\begin{inparaenum}[(1)]
\item Virtual flow control messages are used. They are sent without
  delay in the end of each period. We justify this simplification
  since flow control messages are very sparse.
\item Packets of only one stream are sent along each link in each time slot.
This simplification only reduces the throughput of the implementation.
\end{inparaenum}

\subsection{Scenarios}
We ran the experiments on two main types of arrangements of the nodes
in the plane: a circle and a grid.

\begin{enumerate}
\item In the grid arrangement, we positioned $49$ nodes in a
  $1km\times 1km$ square. The nodes are positioned in a $7\times 7$
  lattice, so that the horizontal and vertical distance between
  adjacent nodes is $1000/7=142$ meters (see Fig.~\ref{fig:scenario
    grid}).    The source and destination of the streams
  in the grid arrangement are chosen randomly.

\item In the circle arrangement, we positioned $24$ nodes on a circle
  of radius $500$ meters.  The nodes were positioned every $360/24$
  degrees. %(see Fig.~\ref{fig:scenario circle}).
  The source and
  destination of the streams in the circle arrangement are chosen
  deterministically as follows: $a_1=\lceil 24/k \rceil$, $b_i=(a_i + \lfloor
  24/k \rfloor) \bmod 24$, $a_{i+1}=b_i$, where $k$ denotes the number of streams.
\end{enumerate}

We point out that random locations of $50$ nodes in a square kilometer
induces a communication graph with a high degree and a diameter of
$2$ or $3$~\cite{marina2010topology}. In addition, the interference set of
each link contains almost all the other links. Hence, this setting has a
low capacity and is not an interesting setting for the problem we
study.

The requests demand $d^*_i$ is set to $10$Mbps. Such a demand with
$k\geq 6$ streams is above the capacity of the network. This enables
us to study the performance in a congested setting.

%\subsection{Scenarios}
%We ran the experiments on a circle with $24$ nodes (evenly spread
%requests) and a grid with $49$ nodes (random requests). We used $k\in\{8,12,16\}$
%requests each with a demand of $10$ Mbps.

\subsection{Benchmarks}
We ran the experiments using six algorithms.
\begin{enumerate}
\item \algA. In the \algA\ benchmark all three parts of our algorithm
  are used: computation of a multicommodity flow with interference
  constraints, the path-peeling scheduler, and the Flow-Control
  algorithm.

\item \algBS. A shortest path maximum bottleneck routing algorithm
  with the path-peeling scheduler.  Let $\pps(e)$ denote the number of
  packets-per-slot in the \MCS\ used by the link $e$.  Let $\hops(p)$
  denote the number of hops along a path $p$.

  We define a (lexicographic) order over paths from $a_s$ to $b_s$ as
  follows: $p \leq q$ if (1) $\min_{e\in p} \pps(e) \geq \min_{e\in q}
  \pps(e)$ or (2) $\min_{e \in p} \pps(e) = \min_{e \in q} \pps(e)$
  and $hops(p) \leq \hops(q)$.  Formally, in \algBS, the stream $s$ is
  routed along a path $p$ that is minimal in the lexicographic order.

  In \algBS, the paths are computed in an oblivious manner, namely,
  congestion does not play a role. This means that we must execute a
  flow control algorithm to adjust the data-rate.

  Each stream in the \algBS\ benchmark is assigned a random
  frequency channel.
\item The remaining algorithms are \algD\ (only multi-commodity flow
  without interference constraints without a scheduler), \algC\ (only
  multicommodity flow with interference constraints without a
  scheduler), \algE\ (multi-commodity flow without interference
  constraints with a scheduler), \algB\ (shortest paths but without a
  scheduler).  A detailed description appears in
  Appendix~\ref{sec:bench}. We point out that whenever the scheduler
  is not invoked, each node must have $3$ WNICS. The reason is that a
  node does not know the frequencies of incoming packets.
\end{enumerate}

We made the following change in the WiFi WNICs when there is a
scheduler. The noise threshold for allowing a transmission of an RTS
frame is reduced to match the interference distance. The reduced
threshold relaxes the conservative collision avoidance to allow for
simultaneous transmissions by links approved by the scheduler.

\subsection{Results}

\paragraph{Comparison between \algA\ and \algBS .}
We focus on two properties: min-throughput (i.e., the lowest throughput
over all the streams) and the end-to-end delay.
\begin{table}
\centering \scriptsize
\begin{tabular}{|| l || c | c | c ||}
\hline
k & \algA 's            & \algBS 's & Ratio\\
    &  min Throughput   & min Throughput & \\
\hline
& Mbps & Mbps & \\
\hline\hline
8 & 0.576 & 0.45  & 1.28  \\
12 & 0.448 & 0.325 & 1.3785 \\
16 & 0.368  & 0.22 & 1.6727 \\
\hline
\end{tabular}
\caption{Comparison of steady state min-throughput between \algA\ and \algBS\ in the
  grid scenario.  The number of requests is denoted by $k$. }
\label{tab:minthroughput}
\end{table}

Table~\ref{tab:minthroughput} lists the effect of the number of
requests $k$  on the minimum throughputs of \algA\ and \algBS\ in the
grid scenario.  \algA\ outperforms \algBS\ by $28$-$67\%$.

\begin{figure}%[h!]
  \centering
  \subfloat[Comparison of throughput]{\label{fig:small11}\includegraphics[width=0.5\textwidth]{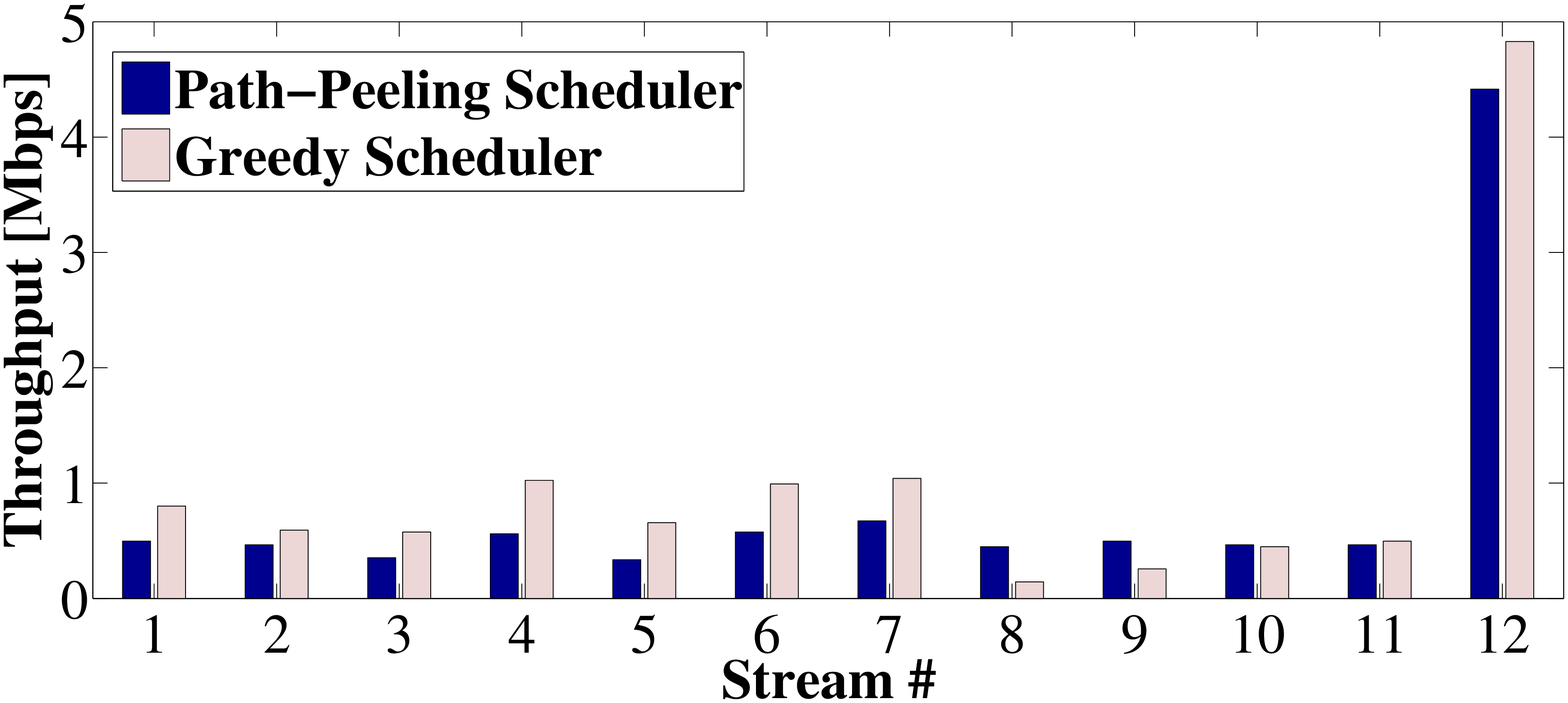}}
  \subfloat[Comparison of end-to-end delay]{\label{fig:small5}\includegraphics[width=0.5\textwidth]{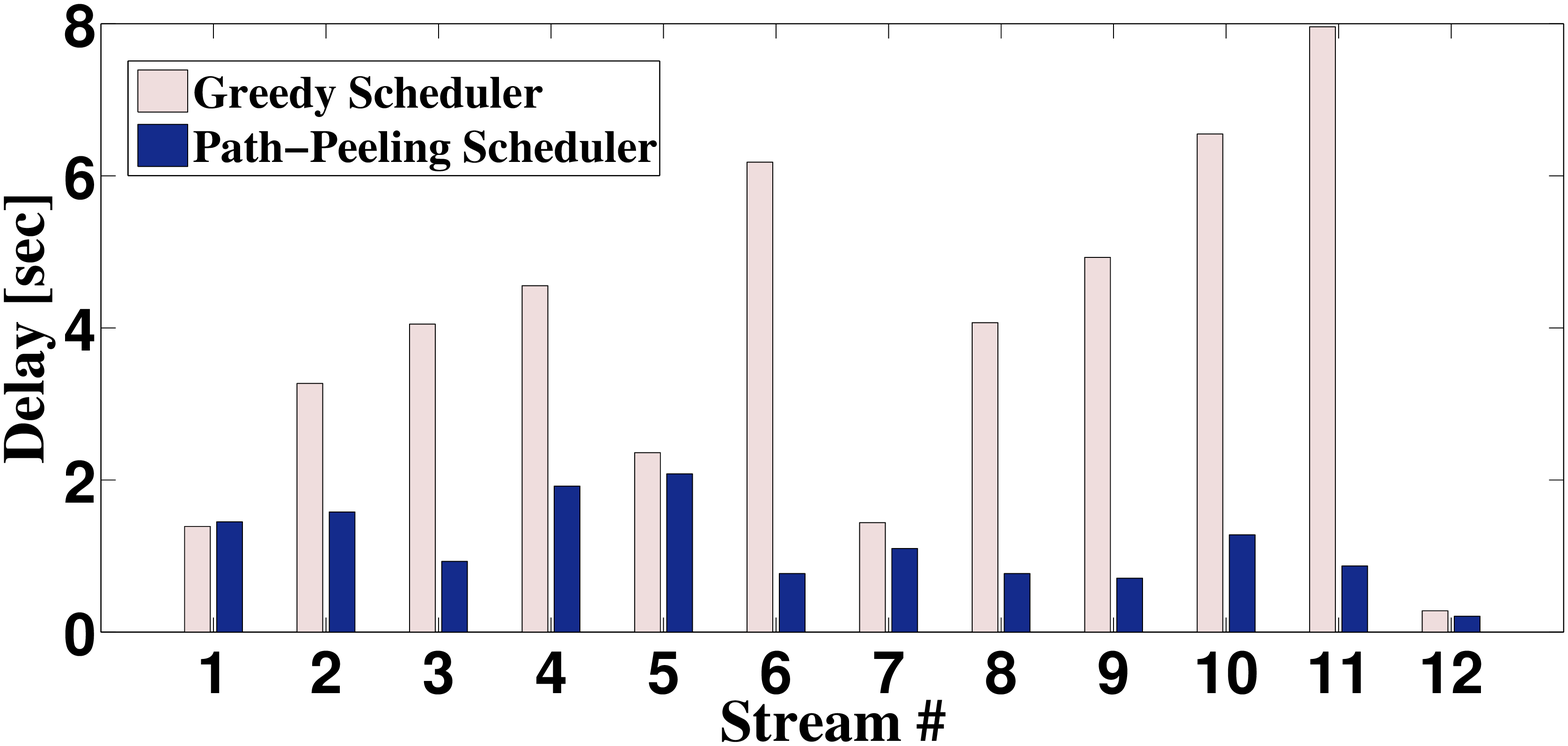}}\\
  \subfloat[Throughput per stream (greedy scheduler)]{\label{fig:small9}\includegraphics[width=0.5\textwidth]{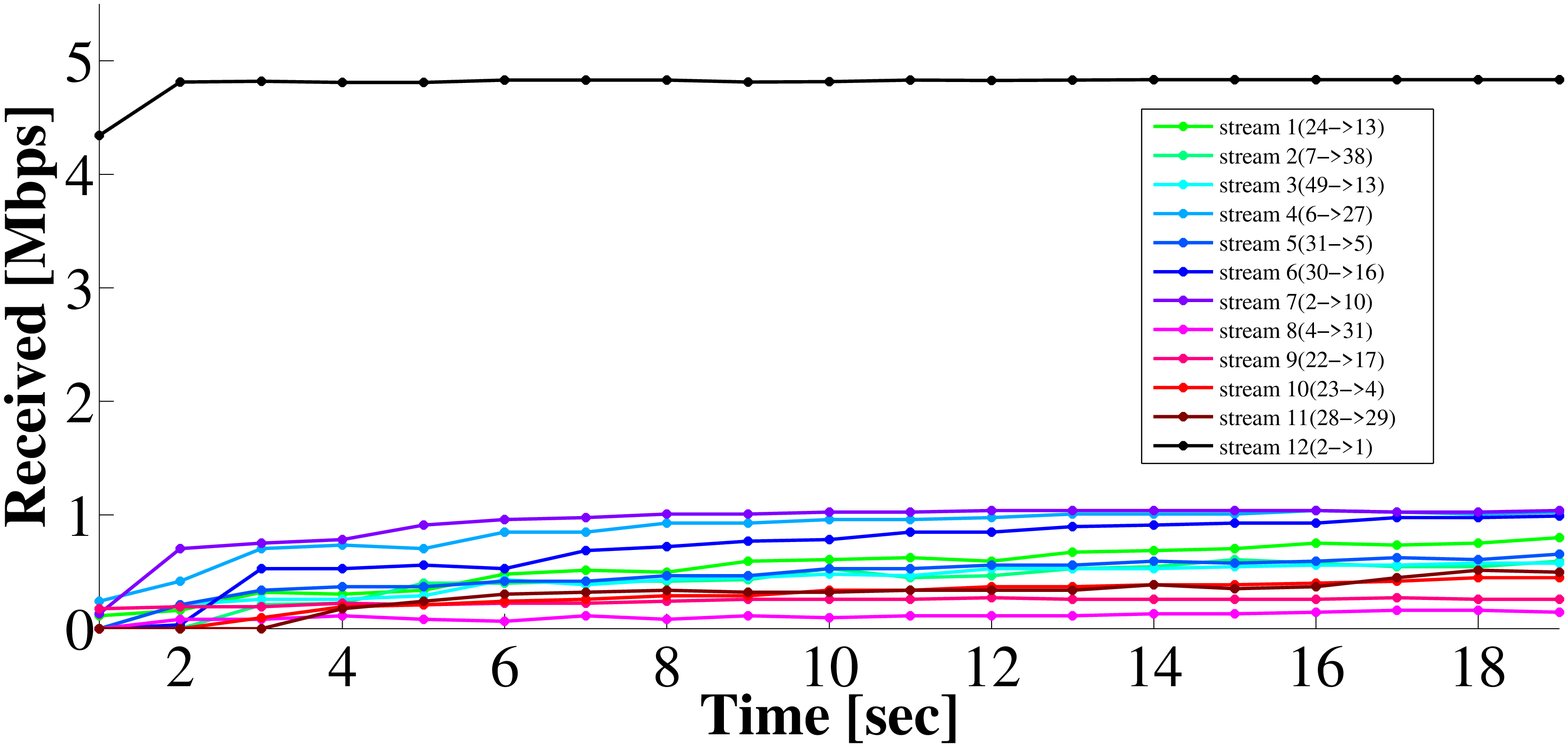}}
  \subfloat[Throughput per stream (path-peeling scheduler)]{\label{fig:small10}\includegraphics[width=0.5\textwidth]{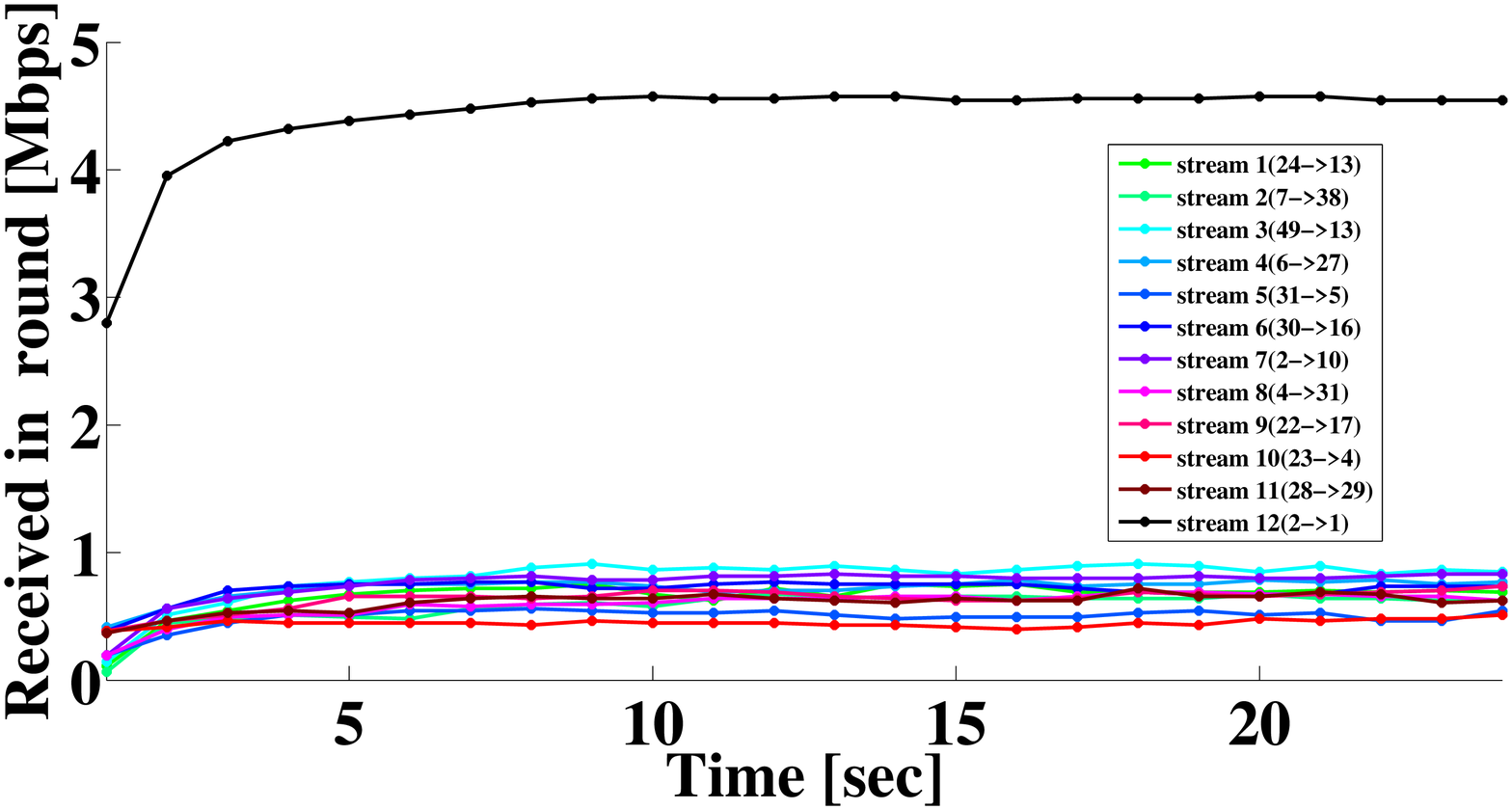}}
\caption{Comparison of \algA\ with the greedy scheduler and the path-peeling scheduler in  the grid arrangement with $k=12$ and $d^*_i=10$Mbps. The experiment's duration is $25$ seconds.}
  \label{fig:big3}
     \end{figure}

\paragraph{Comparison with Greedy Scheduler.}
In Figure%~\ref{fig:sched}
~\ref{fig:big3} we compare \algA\ with the greedy scheduler
and the path-peeling scheduler.  The path peeling scheduler
significantly reduces the end-to-end delay while slightly reducing the
throughput. Note that the min-throughput is bigger with the
path peeling scheduler (i.e., stream \#8), hence, fairness is
improved.

\begin{comment}
\begin{figure}
\centering
\includegraphics[width=0.45\textwidth]{max_delay_greedy_vs_smart.eps}
\includegraphics[width=0.45\textwidth]{tp_sched_comparison.eps}
\caption{Comparison of end-to-end delays and throughputs between the greedy scheduler
  and the path-peeling scheduler.  \algA\ was run in the grid
scenario with $k=12$ requests and uniform demands $d^*_i=10$Mbps. }
  \label{fig:sched}
\end{figure}
\end{comment}

\paragraph{Benchmark Comparison.}
In Tables~\ref{table_grid} and~\ref{table_circle} we summarize the
measured performance of the benchmarks for the grid and circle
scenarios with $k=12$ requests and a demand $d^*_i=10$Mbps for each
stream. The experiment's duration is $25$ seconds.
Notice that the \algA , \algE , and \algBS\ benchmarks use only one WNIC
per node, while the other benchmarks use three WNICs per node.

We begin by discussing the grid scenario:
%\begin{inparaenum}[(1)]
\begin{enumerate}[(1)]
\item It is evident that the dropped packets
    rate is exceedingly high when the scheduler is not used. The reason is
    that the flow control algorithm fails to stabilize the queue length, as
    one would expect.
\item The minimum throughput obtained by \algA\ is the highest.
\item The sum of the throughputs obtained by \algA\ is $35$\% higher
  than that of \algS, and $140$\% higher
  than that of \algBS.
\item \algA\ uses longer paths to avoid congestion and interference.
\item The scheduled benchmarks (i.e., \algA\, \algBS\ and \algE ) have a very small drop rate.
\item  All benchmarks have reasonable \PER.
%\end{inparaenum}
\end{enumerate}

The circle scenario is highly symmetric (i.e., $24$ nodes and $12$
streams) so we can suggest an optimal solution.  This solution is a
table with two slots.  In slot $1$, the ``odd'' links are scheduled.
In slot $2$, the ``even'' links are scheduled.  We need only two
frequencies: so that links separated by a link use different
frequencies.
In this solution, the flow along each link is half its capacity.

Since the capacity of a link in this scenario is $8.2$ Mbps , it follows that
the flow along a link is $4.1$ Mbps. Since streams are routed along
disjoint paths, the throughput per stream is also $4.1$ Mbps.

In the circle scenario we obtained the following results:
\begin{enumerate}[(1)]
%\begin{inparaenum}[(1)]
\item The minimum throughput obtained by \algA\ is $0.49$ that of \algB, and $0.5$ that of \algBS.
\item The sum of the throughputs obtained by \algB\ is only $75$\% more than that of
  \algA.
\item Drop rate are small also for \algB.
%\end{inparaenum}
\end{enumerate}

In light of the fact that \algA\ uses a single WNIC per node, it is
clear that it outperforms all other algorithms. Most importantly, the
end-to-end delay in \algA\ is much shorter.

  \begin{table*}
\centering\scriptsize
\begin{tabular}{|| l || c | c | c | c | c|| c | c || c | c || c | c | c ||}
\hline
& \#radios  & \multicolumn{3}{|c|}{throughput}  & Delay & \multicolumn{4}{|c||}{delay based on hops} & hops & drops & \PER \\
\hline
& per node & min & sum & max & max & \multicolumn{2}{|c||}{max} & \multicolumn{2}{|c||}{min} & avr & max & avr \\
\hline
& & Mbps & Mbps & Mbps & sec & \#hops & sec & \#hops & sec & & \% & \% \\
\hline\hline
\algA & 1 & 0.512 & 12.08 & 4.544  & 2.25 & 14 & 2.25 & 1 & 0.2 & 7.17 & 0.7 & 1.85 \\
\algB & 3 & 0.064 &  8.9  & 1.568  &  \textbf{24}  &  9 &  23  & 1 & 1.35 & 4.5 & 24  & 1.09 \\
\algBS & 1 & 0.402  & 5.02 & 0.4342 & 1.5 & 9  & 1.32  & 1 & 0.95 & 4.5   & 0   & 0.15 \\
\algC & 3 & 0.064 & 4.32  & 1.2 & 3.9 & 14 & 2.6 & 1 & 1.2 & 7.17 & \textbf{95}   & 5.54 \\
\algD & 3 & 0 & 3.14  & 0.704  & 4.9 & 12 & 2.4  & 1 & 1.3  & 6.67  & \textbf{78}   & 6.63 \\
\algE & 1 & 0.1548  & 4.26  & 0.8  & \textbf{19}  & 12 & 15  & 1 & 3.7  & 6.67  &  0.5   & 0.87 \\
\hline
\end{tabular}
\caption{Comparison of the benchmarks for the grid scenario with $k=12$ requests and $d^*_i=10$Mbps for each stream. The experiment's duration is $25$ seconds. }
\label{table_grid}\label{tbl:grid}
%\end{table*}

%\begin{table*}
\centering\scriptsize
\begin{tabular}{|| l || c | c | c | c | c|| c | c || c | c || c | c | c ||}
\hline
& \#radios  & \multicolumn{3}{|c|}{throughput}  & Delay & \multicolumn{4}{|c||}{delay based on hops} & hops & drops & \PER \\
\hline
& per node & min & sum & max & max & \multicolumn{2}{|c||}{max} & \multicolumn{2}{|c||}{min} & avr & max & avr \\
\hline
& & Mbps & Mbps & Mbps & sec & \#hops & sec & \#hops & sec & & \% & \% \\
\hline\hline
\algA & 1 &  2.37 & 32.1  & 2.97 & 1.23 & 2 & 1.23 & 2 & 0.68 & 2 & 0   & 0 \\
\algB & 3 & 4.82 & 56.2 & 4.88  & 2.5  & 2  & 2.5  & 2 & 1  & 2     & 34   & 0 \\
\algBS & 1 & 1.98 & 30.4 & 2.78  & \textbf{5.35}  & 2  & 4.1  & 2 & 0.35  & 2     & 0   & 0 \\
\algC & 3 & 1.35  & 18.4 & 1.75  & 1.5 & 2 & 1.5  & 2 & 1 & 2 & \textbf{36}  & 3 \\
\algD & 3 &  1.3 & 17.7 & 1.7 & 1.7 & 2 & 1.7  & 2 & 1 & 2 & \textbf{14}  & 2.48 \\
\algE & 1 &  0.85 & 15.42 & 1.75 & 2.64  & 2 & 2.64 & 2 & 4  & 2 & 0.65 & 0 \\
\hline
\end{tabular}
\caption{Comparison of the benchmarks for the circle scenario with $k=12$ requests and $d^*_i=10$Mbps for each stream. The experiment's duration is $25$ seconds. }
\label{table_circle}\label{tbl:circle}
\end{table*}

\begin{figure}[H]
      \centering
        \subfloat[Grid scenario layout]{\label{fig:small18}\includegraphics[width=0.5\textwidth, height = 0.4\textheight]{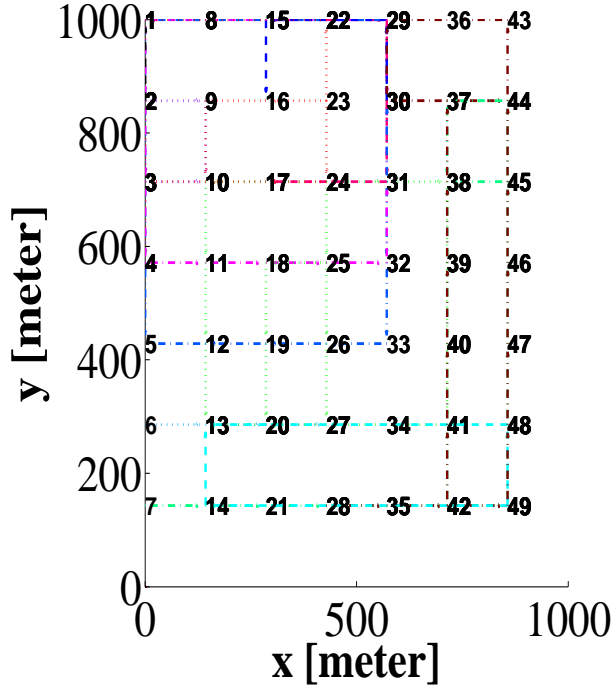}}
        \caption{The grid scenario with $49$ nodes, and $k=12$
          requests.  Flow paths, computed by \algA, are depicted.  An
          example of the splitting of flow can be seen for the request
          from node $49$ to node $13$. This request is split in to
          two paths along the perimeter of the rectangle
          $(13,48,14,49)$.}
      \label{fig:big6}\label{fig:scenario grid}
     \end{figure}

\paragraph{Routing results.}
The routing result for $k=12$ streams is depicted for the grid
scenario in Fig.~\ref{fig:scenario grid}
%in Appendix~\ref{sec:figs}. % and for the circle scenario in Fig.~\ref{fig:scenario circle}.

Note that a request may be split among multiple paths.

\paragraph{Comparison between \algB\ and \algBS.}
Table~\ref{tab:dropsBS} depicts the effect of applying the path-peeling scheduler on the greedy algorithm \algB, that is we compare algorithm \algB\ and \algBS\ in the grid scenario.
Recall that \algBS\ has a single radio in every node.
The number of requests $k$ is 12. The demand for every stream request is $10$ Mbps.

It can be seen that \algBS\ has no drops, hence our algorithm \algA\ is compared to \algBS.
\begin{table}
\centering\small
\begin{tabular}{|| l || c | c ||}

\hline
 & \algB 's            & \algBS 's \\
    &  drops percentage   & drops percentage \\
\hline\hline
max     & 46 & 0  \\
min     & 0 & 0 \\
average & 11.66  & 0 \\
\hline
\end{tabular}
\caption{
  Comparison of the drop percentage between \algB\ and \algBS\ in the grid scenario. The drop percentage is
  the ratio between the number of
  dropped packets and the number of the transmitted packets. The simulations used $k=12$ streams and a uniform demand of $d_i^*=10$ Mbps.}
\label{tab:dropsBS}
\end{table}

\paragraph{Effect of number of streams.}
Table~\ref{tab:minthroughput} depicts the effect of the number of
requests $k$  on the minimum throughputs of \algA\ and \algBS\ in the
grid scenario.  The number of requests $k$ is $8,12,16$. The demand
for every stream request is $10$ Mbps.

Clearly, the min-throughput decreases as the number of requests
increases, as the same resources need to serve more requests.
The advantage of \algA\ is maintained for this range of requests.

\begin{figure}%[h!]
      \centering
        \subfloat[Three radios compared to one radio]{\label{fig:small1}\label{fig:radio13}\includegraphics[width=0.5\textwidth]{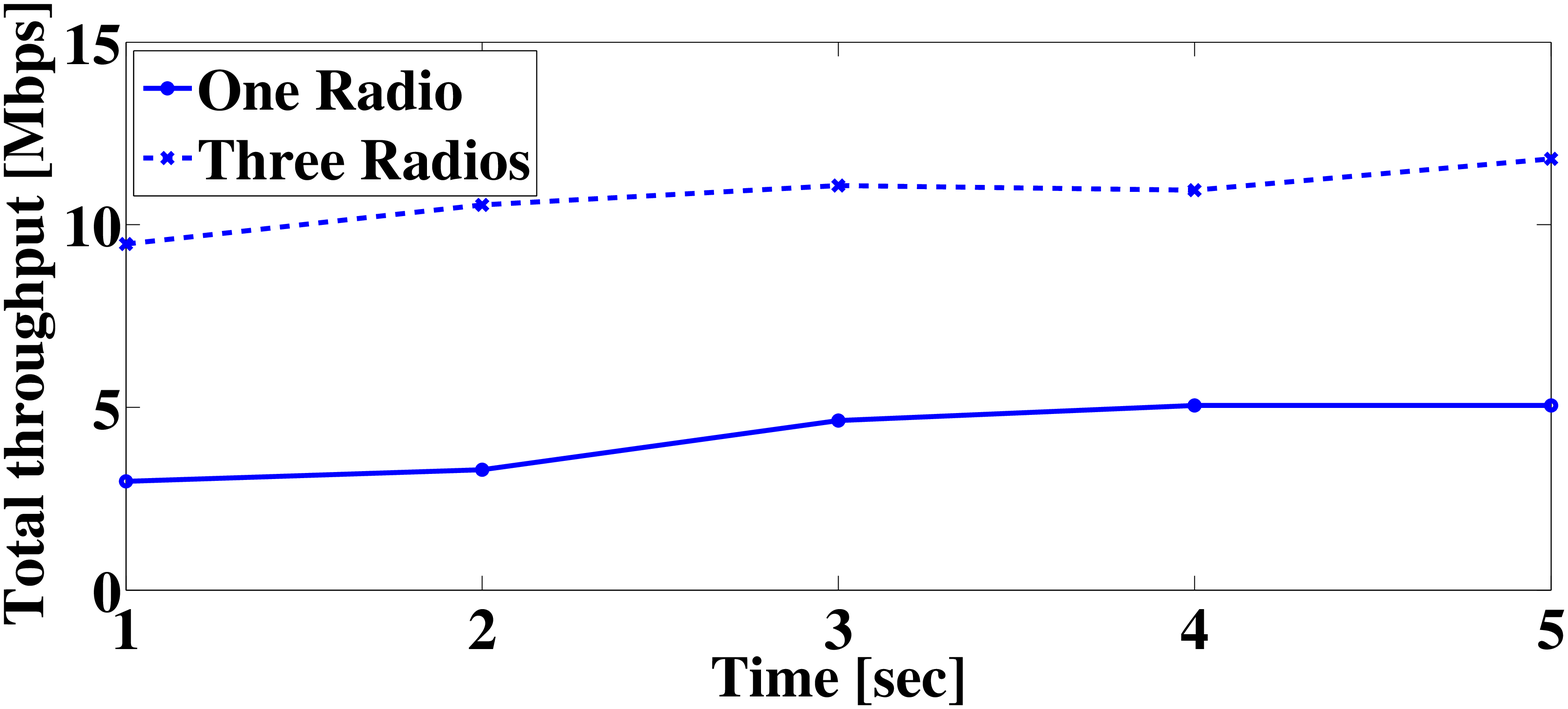}}
       \subfloat[Fairness comparison for $16$ requests]{\label{fig:small3}\label{fig:fair}\includegraphics[width=0.5\textwidth]{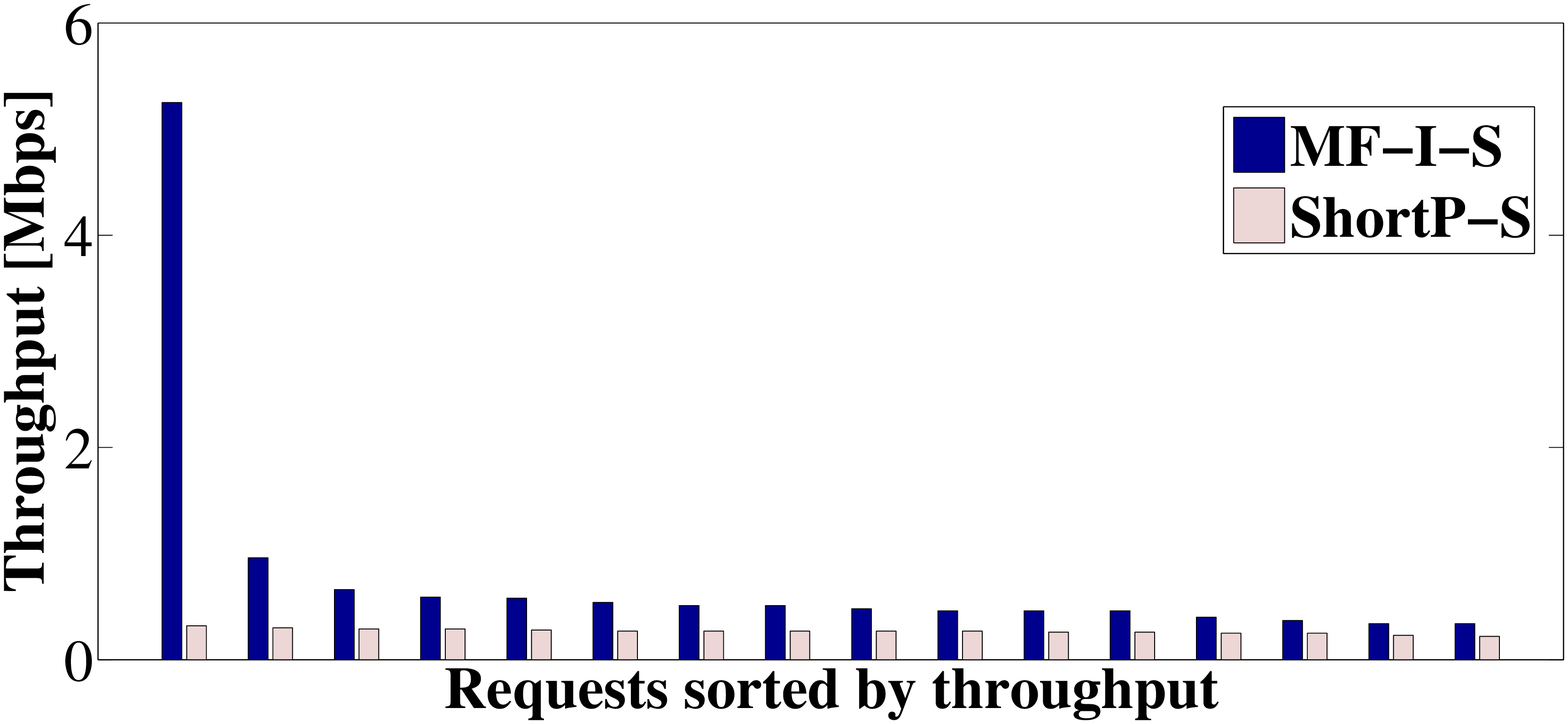}}
          \caption{ (a)~Comparison of total throughput between one and three radios, for \algS\ in the grid
          scenario with $k=12$ and $d^*_i=10$Mbps. (b)~ Throughput comparison between \algBS\ and \algA. Streams in each benchmark are sorted by their
          throughput for the grid arrangement with $k=16$. The experiment's duration is $5$ seconds. }
     \end{figure}

\paragraph{Effect of single radio on \algS.}
Since \algS\ uses $3$ WNICs per node, we experimented also with a single WNIC per node.
Figure~\ref{fig:radio13} depicts the effect of single radio on total throughput of \algS.
The ratio is almost constant and equals $3$, as expected.

The experiment was made in the grid arrangement, where The number of requests is $k=12$.
The demand for every stream request is $10$ Mbps.

\paragraph{Fairness.}
Figure~\ref{fig:fair} compares the sorted throughputs achieved for the
streams in \algA\ and \algBS.  The experiment was made in the grid
arrangement, with $k=16$ streams and $d^*_i=10$ Mbps.

%Although \algBS\ has a higher total throughput, the $7$ with the
%smallest throughputs have less throughput in \algBS\ than in \algA.
%This means that \algBS\ favors short links, and is less fair than \algA.
It can be seen that every request achieves less throughput in \algBS\ than in \algA.
Note that \algBS\ balances the throughput among its requests.

\paragraph{The Flow-Control algorithm.}

\begin{figure}%[h!]
      \centering
        \subfloat[Flow control rate]{\label{fig:fcsmall1}\includegraphics[width=0.5\textwidth]{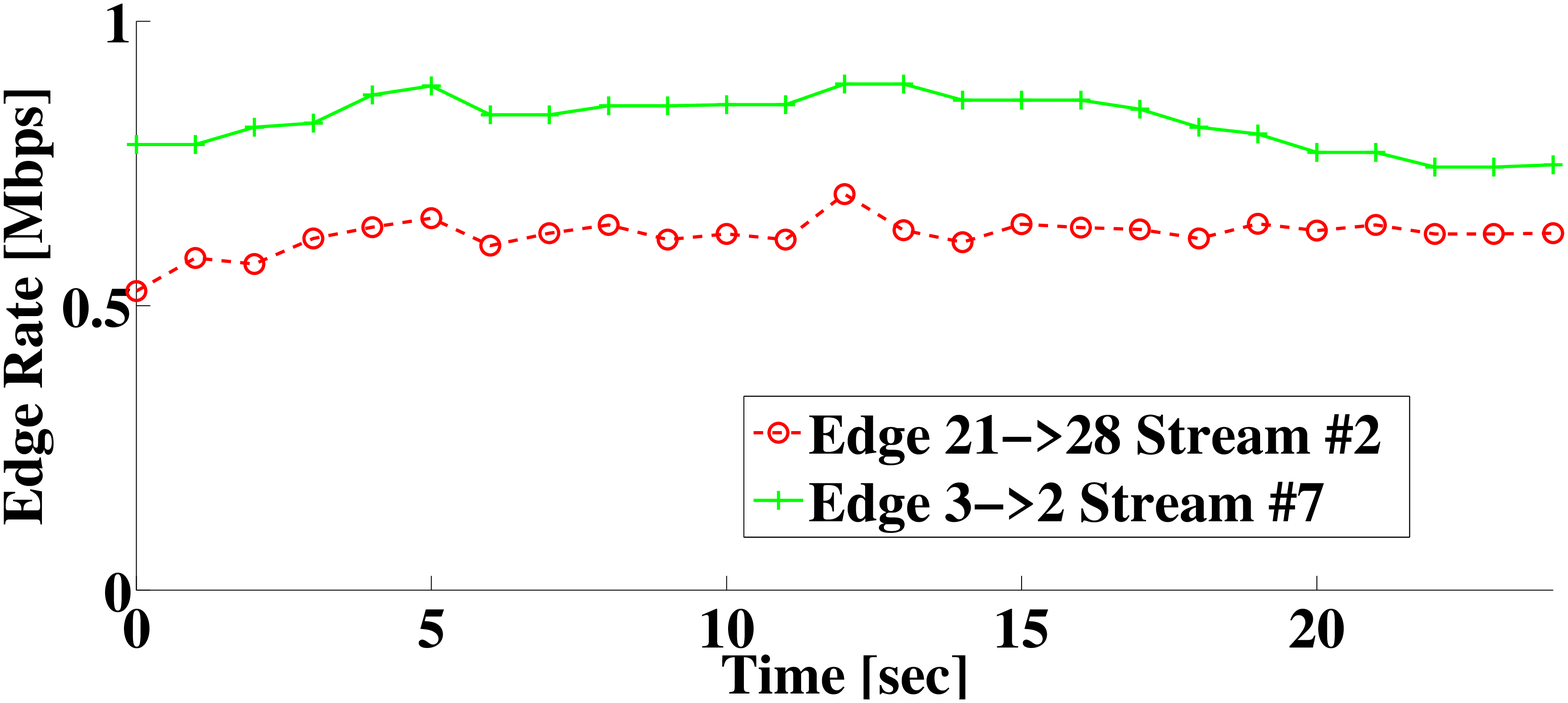}}
        \subfloat[Queue length max stream \#10]{\label{fig:fcsmall2}\includegraphics[width=0.5\textwidth]{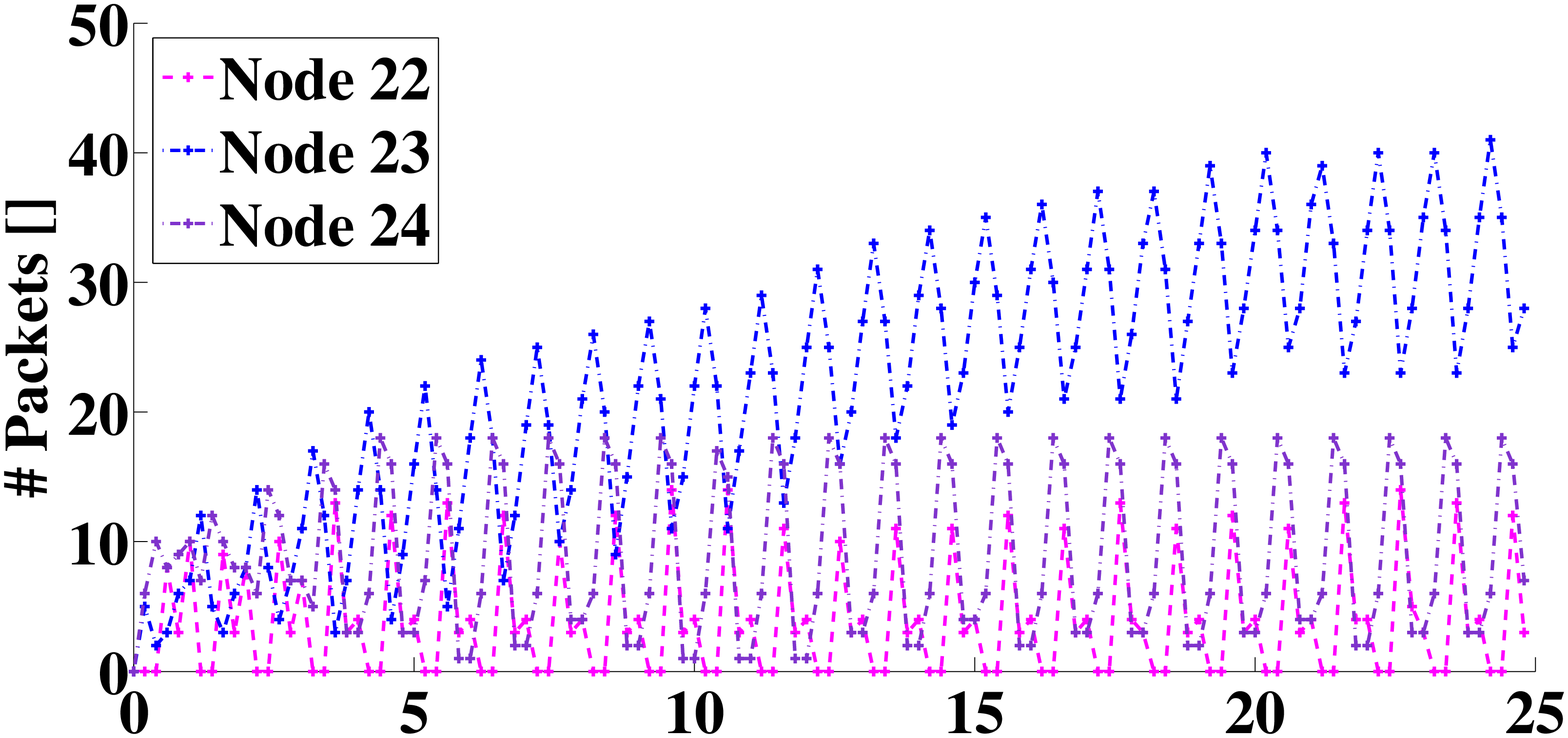}}\\
        \subfloat[Maximum packet drop over time (stream \#1)]{\label{fig:fcsmall3}\includegraphics[width=0.5\textwidth]{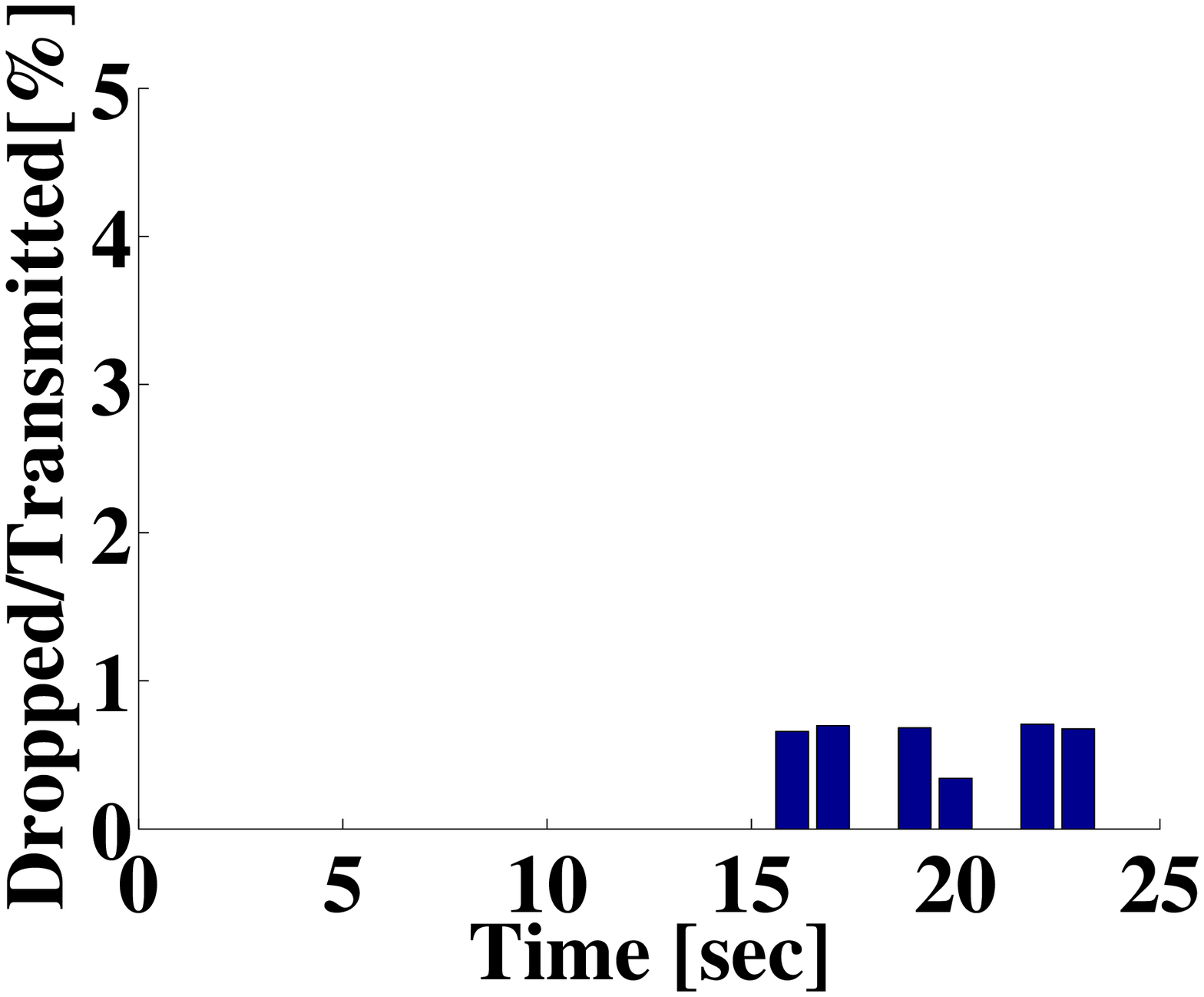}}
        \subfloat[Throughput stability comparison]{\label{fig:fcsmall4}\includegraphics[width=0.5\textwidth]{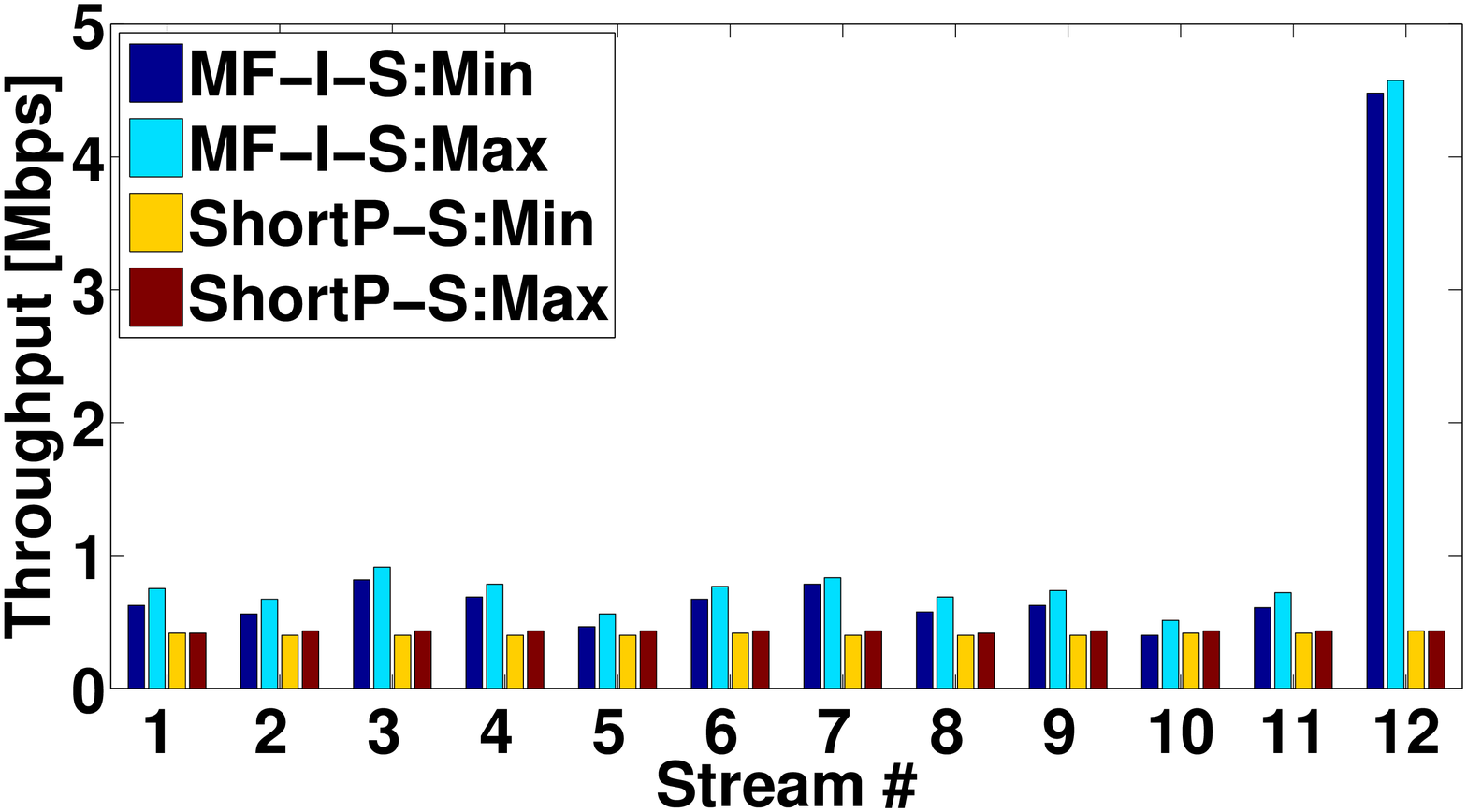}}\\
        \caption{Influence of the Flow-Control algorithm in  the grid arrangement with $k=12$ and $d^*_i=10$Mbps. The experiment's duration is $25$ seconds: (a)~Change in the requested packet rate by the flow-control over time in \algA , (b)~Queue lengths of a stream in three different nodes in \algA , (c)~Ratio of dropped packets to transmitted packets over time in worst stream in \algA , (d)~Comparison of stability of throughput between \algA\ and \algBS.
        }
    \label{fig:fcbig}
     \end{figure}

Figure~\ref{fig:fcbig} depicts the effects of the Flow-Control
algorithm in the grid arrangement with $k=12$ and $d^*_i=10$Mbps.  In
Fig.~\ref{fig:fcsmall1}, the requested rates $R(e,s)$ are depicted. It
can be seen that only slight perturbations occur over time. This
justifies our simplified implementation that uses virtual flow-control messages.

In Fig.~\ref{fig:fcsmall2}, the queue lengths of the stream in three
different nodes are depicted.  The oscillation is due to the periodic
schedule. The queue length is controlled and stabilizes.

In Fig.~\ref{fig:fcsmall3}, the drop ratio is depicted for the worst
stream. The drop rate ranges from $0$ to $1$\%.

In Fig.~\ref{fig:fcsmall4}, the differences between the maximum and
minimum throughput for \algA\ and \algBS\ are depicted for all streams
two seconds after the beginning of the experiment.
It is evident that \algA\ and \algBS\ are stable since the differences are smaller.

%\begin{comment}
%  \textbf{suggest other ways to implement the flow-control messages
%    (incorporate in ACKs, send in the beginning of a time slot,
%    reserve control slots).}
%\end{comment}

\section{Conclusions}
The algorithm consists of two parts: a multi-commodity flow
computation and a scheduler.  Our simulations demonstrate the
robustness of the scheduler.  Namely, the flows $\mf$ that are
supported by the time-slotted frequency table $A$ are successfully
routed in the \SINR-model. Thus, in our simulations the modified graph
model results with \SINR-feasible schedules.

The role of the multi-commodity flow computation with interference
constraints is to maximize the minimum throughput. Indeed, in the grid
scenario, routing along shortest paths resulted with smaller
throughputs.

The flow control algorithm succeeds in stabilizing the queue lengths
for all benchmarks that used the scheduler. Without the scheduler,
stability was not obtained, and many packets were dropped.

Our results show that one can compute a routing and scheduling that
succeeds in the \SINR-model while using a simpler interference model.
In addition, we successfully combined the various goals required to
support video streaming.

\section{Discussion}
We propose a centralized algorithm for computing a routing,
scheduling, and frequency assignment for real-time video streams in
static ad-hoc wireless networks.  The algorithm consists of two parts:
a linear program and a scheduler.  In addition, each node locally runs
a flow-control algorithm to control the queues and stabilize data-rate
along the links.  Although the algorithm is centralized, it can be
executed by multiple nodes in the network provided that they hold full
information of the network (i.e., locations, requests).  The output of
the algorithm consists two tables that can be easily broadcast to all
the nodes.

We implemented the algorithm and experimented using a setting that
uses the physical model (with a 802.11g MAC) to verify the validity of
the algorithm.  Our experiments show that the traffic routed and
scheduled by the algorithm is successfully delivered in two congested
scenarios in the \SINR-model.

We propose a scheduling algorithm, called the path peeling scheduler,
that is designed to reduce the end-to-end delay incurred by the greedy
scheduler. The path peeling scheduler succeeded in reducing the delay
in streams with many hops. Even in a congested scenario, the path
peeling scheduler successfully scheduled at least $70\%$ of the flow.

\ifnum\conff=0
  \subsection*{Acknowledgments}
  We thank Nissim Halabi and Magnus M. Halldorsson for useful discussions.
This work was supported in part by the Israeli Ministry of Industry and Trade under project MAGNET by the RESCUE Consortium.
\fi
%

%\nocite{*}
\bibliographystyle{alpha}
\bibliography{wireless}

\newpage
\appendix

\section{TX-RX Algorithm} \label{sec:algs}
A listing of the TX-RX algorithm appears as Algorithm~\ref{alg:tx-rx}.
We elaborate below how a queue $Q(v,s)$ with the highest priority for
transmission along link $e$ in line~\ref{line:priority} in the
Transmit procedure is defined.

Upon invocation of Transmit$(e,j)$, where $e=(v,u,m)$, the node $v$
needs to decide which packet to transmit. The node $v$ uses the
multi-flow table $\mf$ to determine the set $S_e$ of streams that are
routed along $e$.  Since delay is a major issue, it is reasonable to
use an EDD-like policy (Earliest Due Date), i.e., pick the oldest packet in the queues
$Q(v,s)$, for $s\in S_e$.  However, such a policy ignores the
remaining number of hops a packet needs to traverse. We prefer the
approach that emphasizes fairness. That is, assign a priority that
equals the ratio of the number of packets of stream $s$ transmitted
along $e$ in the last period divided by the required number. The lower
this ratio, the higher priority of the stream. This approach also
combines well with the flow control algorithm described in
Sec.~\ref{sec:flow control}.

\begin{algorithm}[H]
  \caption{TX-RX$(v)$ - a local transmit-receive algorithm for node
    $v$ as specified by a time-slotted frequency table $A$.}
\label{alg:tx-rx}
For time slot $t'=0$ to $\infty$ do
  \begin{enumerate}%1
\item $t = t' \pmod T$.
  \item if $\Ein(v)\cap A[\cdot,t ]\neq \emptyset$ then
    \{reception mode\}
    \begin{enumerate}%2
    \item Let $e\in \Ein(v)\cap A[\cdot,t]$, where $e(u,v,m)\in
      A[j,t]$.
    \item While slot $t$ is not over call Receive$(e,j)$.
    \end{enumerate}%2
  \item if $\Eout(v)\cap A[\cdot,t ]\neq \emptyset$ then
    \{transmission mode\}
    \begin{enumerate}%2
    \item Let $e\in \Eout(v)\cap A[\cdot,t ]$, where
      $e(v,u,m)\in A[j,t]$.
    \item While slot $t$ is not over call Transmit$(e,j)$.
    \end{enumerate}%2
  \end{enumerate}%1
Receive$(e,j)$ - where link $e=(u,v,m)$ and $j$ is a frequency channel.
\begin{enumerate}
\item Set tuner to reception in frequency channel $j$.
\item Upon reception of a packet $p$ from stream $s$, insert $p$ to $Q(v,s)$.
\end{enumerate}

Transmit$(e,j)$-  where link $e=(v,u,m)$ and $j$ is a frequency channel.
\begin{enumerate}
\item Set tuner to transmission in frequency channel $j$.
\item\label{line:priority} Pick a queue $Q(v,s)$ with a highest priority for transmission along $e$.
\item $p \gets DEQUEUE(Q(v,s))$.
\item Transmit $p$ along $e$.
\end{enumerate}
\end{algorithm}

\section{The Linear Programming Formulation}\label{sec:LP}
\begin{figure*}
  \begin{align}
    \max ~~~\rho + \lambda \cdot\sum_{i=1}^k d^*_i \cdot \rho_i~~~~~~&\text{subject to}&\\
    f^j_i(e) &\geq 0&\forall i\in [1..k], \forall j\in [1..3], \forall e\in E \label{eq:pos}\\
    \sum_{j=1}^3 f_i^j(e) &=f_i(e)& \forall e\in E\label{eq:freq1}\\
    \sum_{i=1}^k f_i^j(e) &=f^j(e)& \forall e\in E\label{eq:freq2}\\
    \sum_{e \in \Eout (v)} f_i(e) - \sum_{e \in \Ein (v)} f_i(e) &=0 &\forall i\in [1..k],\forall v\in V\setminus \{a_i,b_i\}\label{eq:conservation}\\
    \sum_{i=1}^k f_i(e) &\leq c(e) &\forall e\in E \label{eq:cap}\\
    \sum_{e \in \Eout (v)} f_i(a_i) - \sum_{e \in \Ein (v)} f_i(a_i)&=d^*_i \cdot \rho_i &\forall i\in [1..k]\label{eq:rho_i}\\
    \rho &\leq \rho_i&\forall i\in [1..k]\label{eq:rho}\\
    \frac{f^j(e)}{c(e)} + \sum_{j'< j} \sum_{e'\in E(u)\cup E(v)}
    \frac{f^{j'}(e')}{c(e')} + \sum_{e' \in I_{e}}
    \frac{f^j(e')}{c(e')} & \leq 1 & \forall e=(u,v,m)\in E,\forall
    j\in[1..3] \label{eq:conf}
  \end{align}
\end{figure*}

The main variables of the LP are the flow variables $f_i^j(e)$ which
signify the amount of flow along link $e$ in frequency channel $j$
for stream $i$.  In Eq.~\ref{eq:pos} we require that the flows are
nonnegative.  In Eq.~\ref{eq:freq1} we define $f_i(e)$ to be the
combined flow along $e$ for stream $i$ over all frequency channels.  In
Eq.~\ref{eq:freq2} we define $f^j(e)$ to be the combined flow along
$e$ in frequency channel $j$ over all $k$ streams.
Eq.~\ref{eq:conservation} is simply a flow conservation constraint for
stream $i$ in every intermediate node.  Eq.~\ref{eq:cap} is simply a
capacity constraint for every link.  In Eq~\ref{eq:rho_i}, the supply
ratio $\rho_i$ is defined to be the fraction of the demand for stream $i$
that is supplied.  In Eq.~\ref{eq:rho}, $\rho$ is defined to equal the
minimum supply ratio, i.e., $\rho=\min_i \rho_i$.  Finally, in
Eq.~\ref{eq:conf} the interference constraints are defined; we
elaborate on them below.

The objective is to maximize the minimum supply ratio $\rho$.  As a
secondary objective, we maximize the sum of flows.
Therefore, the constant $\lambda$ in the objective function is small
(e.g., $\lambda=1/20)$.

We point out that the capacity constraints in Eq.~\ref{eq:cap} are
redundant since they are implied by the interference constraints in
Eq.~\ref{eq:conf}.

In our experiments, we noticed that the LP-solver found a solution
with flow cycles. We removed these cycles before applying the
scheduling step.  Interestingly, the issue of flow cycles was not
mentioned in previous
works~\cite{alicherry2005joint11,buragohain2007improved}

\begin{comment}
\section{The Grid Scenario}\label{sec:figs}
\begin{figure}[H]
      \centering
        \subfloat[Grid scenario layout]{\label{fig:small18}\includegraphics[width=0.5\textwidth, height = 0.4\textheight]{scenario_main_grid_12r.eps}}
        \caption{The grid scenario with $49$ nodes, and $k=12$
          requests.  Flow paths, computed by \algA, are depicted.  An
          example of the splitting of flow can be seen for the request
          from node $49$ to node $13$. This request is split in to
          two paths along the perimeter of the rectangle
          $(13,48,14,49)$.}
      \label{fig:big6}\label{fig:scenario grid}
     \end{figure}
\end{comment}

\section{Experimental Results}\label{sec:exres}
\subsection{Benchmarks}\label{sec:bench}
We ran the experiments using six algorithms:
\begin{enumerate}
\item \algS - a shortest path maximum bottleneck routing algorithm.
  Let $\pps(e)$ denote the number of packets-per-slot in the
  \MCS\ used by the link $e$.  Let $\hops(p)$ denote the number of
  hops along a path $p$. In \algS, the stream $s$ is routed along a
  path $p$ from $a_s$ to $b_s$ such that, for every path $p'$ from
  $a_s$ to $b_s$, the following holds:
  \begin{align*}
    \min_{e \in p} \pps(e) &\geq \min_{e \in p'} \pps(e) \text{ , and}\\
    \min_{e \in p} \pps(e) &= \min_{e \in p'} \pps(e)  \\
\Longrightarrow &\hops(p) \leq \hops(p').
  \end{align*}
  The paths assigned to the $k$ streams are divided evenly among the
  three frequency channels.

Each node in this benchmark contains three
  radios. This means that each node contains three standard 802.11g
  WNICs, each working in different frequency channel.  Since the
  frequency channels are non-overlapping, one WNIC may receive while
  another WNIC is transmitting.  Each WNIC receives and transmits
  packets according to the WiFi MAC.  Fairness between the streams is
  obtained as follows.  Each WNIC is given FIFO-queue for each stream,
  the packets of which it needs to transmit. Each WNIC uses a simple
  round-robin policy for determining the queue from which the next
  packet is transmitted.

  The paths are computed in an oblivious manner, namely, congestion
  does not play a role. This means that we must execute a flow control
  algorithm to adjust the data-rate.  To execute the Flow-Control
  algorithm without any changes, we trivially cast this routing to our
  setting as follows.  We define the multi-flow $\mf(e,s)$ to equal
  $d^*_s$ if $e$ is in the path assigned to stream $s$, and $0$
  otherwise. The time-slotted frequency table $A$ has a single time
  slot (i.e., $T=1$) whose duration is one second. Namely, the table
  $A$ has three entries, one for each frequency channel.  The table
  entry for frequency channel $j$ lists the links that use frequency
  channel $j$.
\item \algA. In the \algA\ benchmark all three parts of our algorithm
  are used: computation of a multicommodity flow with interference
  constraints, the path-peeling scheduler, and the Flow-Control
  algorithm,

  We emphasize that in this benchmark, each node contains a single
  radio; namely, each node has a single standard 802.11g WNIC capable
  of hopping between the three frequency channels in the beginning of
  each time slot.

\item \algBS - same as \algB\ except that every node contains a single
  radio. In addition, the path-peeling scheduler is applied, hence
  $T=200$, and a random frequency channel is assigned to every
  request.

\item \algE\ - same as \algA\ except that the LP does not include
  interference constraints. The scheduler resolves interferences, so
  it is interesting to see how much throughput is scheduled by the
  scheduler, and whether this throughput is routed in the simulation.

  We point out that the interference constraints constitute a large
  part of the LP constraints. By omitting them, the LP becomes
  shorter, easier to solves, and naturally, the LP solution has a
  higher throughput.

  Since the LP lacks interference constraints, the scheduler may fail
  to schedule the flow. We modified the scheduler in this case so that
  it augments the table $A$ by adding time-slots. This augmentation
  has an adverse effect of reducing throughput and increasing delay.

\item \algC\ - same as \algA\ but without the scheduler.  Instead, the
  multi-flows are assigned in a single-slot schedule, as in Algorithm
  \algS. We point out that in this benchmark, each node is equipped
  with three WNICs as in Algorithm \algS.

  The motivation for this benchmark is that the multi-flow takes into
  account congestion and interference. Since the WiFi MAC deals with
  avoiding collisions, so it is interesting to see how it succeeds in
  scheduling the multi-flows in a distributed manner.

As in \algS, each node has three WNICS in this benchmark.

\item \algD\ - similar \algS\ except that the streams are routed
  according to a multi-commodity flow. The multi-commodity flow is
  computed by an LP without interference constraints.  This benchmark
  helps understand whether non-oblivious congestion aware routing
  improves performance.

As in \algS, each node has three WNICS in this benchmark.
\end{enumerate}

\end{document}